\documentclass[journal,onecolumn]{IEEEtran}
\usepackage{amsmath}
\usepackage{amssymb}
\usepackage{graphicx}
\usepackage{enumerate}
\usepackage{comment}
\usepackage{mathtools}
\usepackage{scrextend} %for \begin{addmargin}

\newtheorem{thm}{Theorem}
\newtheorem{lemma}[thm]{Lemma}

\newtheorem{proposition}[thm]{Proposition}
\newtheorem{corollary}[thm]{Corollary}

\newcommand\numberthis{\addtocounter{equation}{1}\tag{\theequation}} % To number an equation inside align*
\begin{document}

%\IEEEoverridecommandlockouts
%%%%%%%%%%%%%%%%%%%%%%%%%%%%%%%%%%%%%%%%%%%%%%%%%%%%%%%%%%%%%%%%%%%%%%
%                         THE TITLEPAGE                              %
%%%%%%%%%%%%%%%%%%%%%%%%%%%%%%%%%%%%%%%%%%%%%%%%%%%%%%%%%%%%%%%%%%%%%%

\title{Neural dissimilarity indices that predict oddball detection in behaviour}

\author{\IEEEauthorblockN{Nidhin Koshy Vaidhiyan}, \IEEEauthorblockN{S. P. Arun}, \IEEEauthorblockN{Rajesh Sundaresan} \thanks{This work was supported in part by the Department of Science and Technology. The neuronal and behavioral data used in this study was collected by one of the authors (S. P. Arun) while he was at the laboratory of Prof. Carl Olson, Carnegie Mellon University. The material in this paper was presented in part at the 2012 IEEE International Symposium on Information Theory, Cambridge, MA, USA, July 2012, and was presented in part at the 2015 Information Theory and Applications Workshop, San Diego, CA, USA, February 2015.}}

\maketitle
%%%%%%%%%%%%%%%%%%%%%%%%%%%%%%%%%%%%%%%%%%%%%%%%%%%%%%%%%%%%%%%%%%%%%%
%                              ABSTRACT                              %
%%%%%%%%%%%%%%%%%%%%%%%%%%%%%%%%%%%%%%%%%%%%%%%%%%%%%%%%%%%%%%%%%%%%%%

\begin{abstract}
  Neuroscientists have recently shown that images that are difficult to find in visual search elicit similar patterns of firing across a population of recorded neurons. The $L^{1}$ distance between firing rate vectors associated with two images was strongly correlated with the inverse of decision time in behaviour. But why should decision times be correlated with $L^{1}$ distance? What is the decision-theoretic basis? In our decision theoretic formulation, we modeled visual search as an active sequential hypothesis testing problem with switching costs. Our analysis suggests an appropriate neuronal dissimilarity index which correlates equally strongly with the inverse of decision time as the $L^{1}$ distance. We also consider a number of other possibilities such as the relative entropy (Kullback-Leibler divergence) and the Chernoff entropy of the firing rate distributions. A more stringent test of equality of means, which would have provided a strong backing for our modeling fails for our proposed as well as the other already discussed dissimilarity indices. However, test statistics from the equality of means test, when used to rank the indices in terms of their ability to explain the observed results, places our proposed dissimilarity index at the top followed by relative entropy, Chernoff entropy and the $L^{1}$ indices. Computations of the different indices requires an estimate of the relative entropy between two Poisson point processes. An estimator is developed and is shown to have near unbiased performance for almost all operating regions.
\end{abstract}

\section{Introduction}
\label{Introduction}
We invite the reader to participate in the following visual search tasks. There are two search tasks on page \pageref{fig:fig1a}. Find the oddball image in each of the two configurations. Based on the time taken for each of the tasks, identify which of the two is easier.

Among the two search tasks on page \pageref{fig:fig1a}, most subjects find Task 1 the easier, and Task 2 the tougher. Visual search performance, as measured by the time taken to find the oddball image, should depend on the ``similarity'' of the two images. One has the natural hypothesis:
\begin{quotation}
 $(H)$ The more ``dissimilar'' the two images, the shorter the time taken to find the oddball image.
\end{quotation}

To test such a hypothesis, one needs a quantification of the notion of ``dissimilarity'' between two images. Sripati and  Olson \cite{ref:201001JNS_SriOls} proposed one such measure based on neuronal responses (to the images) in the inferotemporal (IT) cortex of the macaque brain. They conducted experiments to 1) find the time taken by human subjects in visual search for a number of image pairs, and 2) record neuronal responses to the same images from the monkey IT cortex. They found quantitative evidence in support of $(H)$ based on their notion of dissimilarity. We now describe their experiments and recall their findings to set the stage for this paper.

\begin{figure}[t!]
\centering
\includegraphics[scale=0.55]{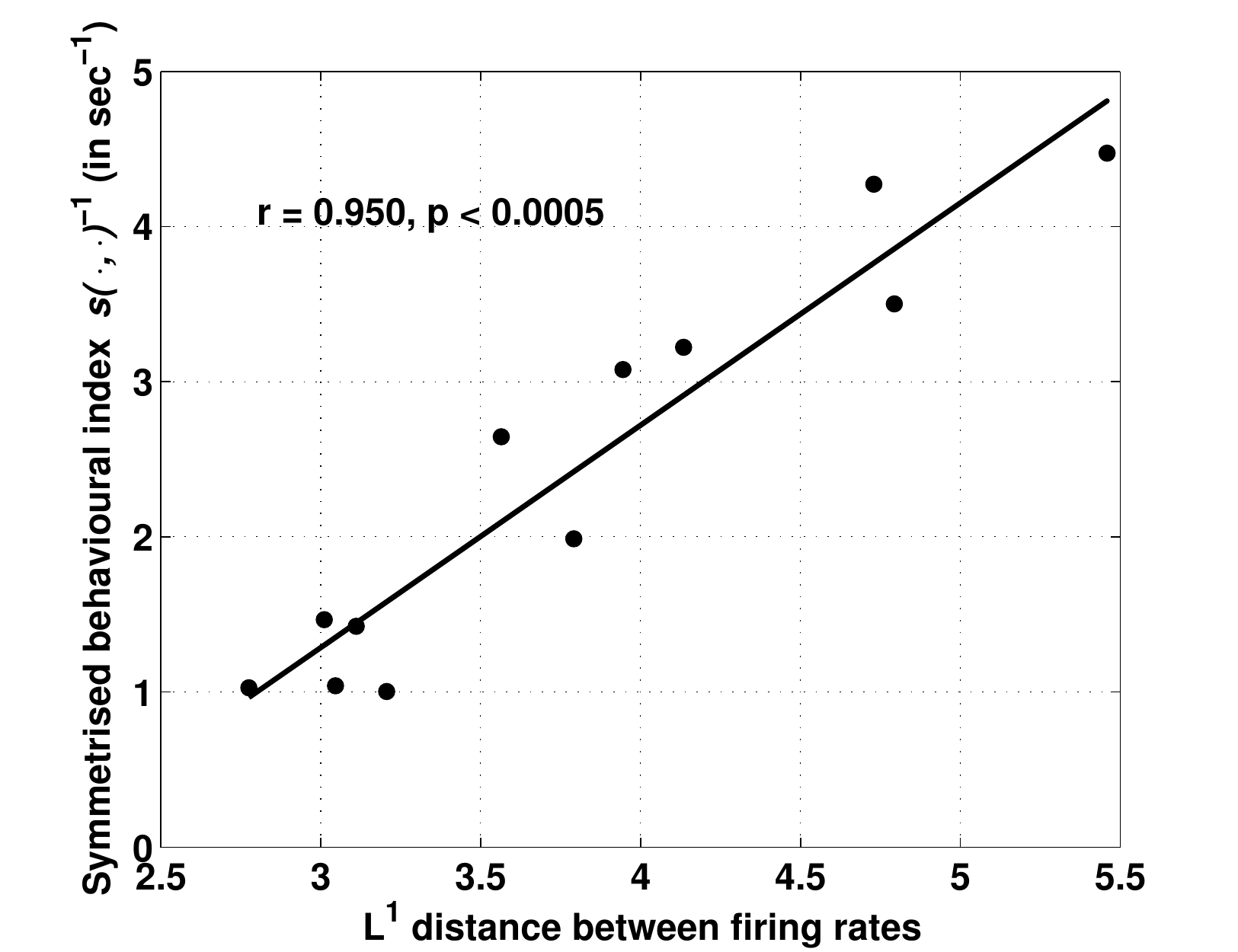}
\caption{Scatter plot of $(s(k,l)^{-1}, \Vert \mathbf{R}_{k} - \mathbf{R}_{l}\Vert _{1})$. Sripati and Olson \cite{ref:201001JNS_SriOls} observed a high correlation of $0.95$ between the inverse of reaction time and their proposed $L^{1}$ distance between the neuronal firing vectors.}
\label{fig:L1_vs_Behavioural}
\end{figure}

The experiments of Sripati and Olson \cite{ref:201001JNS_SriOls} were the following.
\begin{enumerate}
 \item  Six human subjects were shown a picture as in Figure \ref{fig:fig1a}  on page \pageref{fig:fig1a}. Six images were placed at the vertices of a regular hexagon, with one image being different from the others. To be specific, let $I_{k}$ and $I_{l}$ be two images. One of these two was picked randomly with equal probability and was placed at one of the six locations randomly, again with equal probability. The other image was placed in the remaining five locations. The subjects were required to identify the correct half (left or right) of the plane where the oddball image was located. The subjects were advised to indicate their decision ``as quickly as possible without guessing" \cite{ref:201001JNS_SriOls}. The time taken to make a decision\footnote{A baseline motor reaction time for each subject was also estimated in a separate experiment and subtracted to get an estimate of the time to {\it make} a decision. See \cite{ref:201001JNS_SriOls} for details.} after the onset of the image was recorded. This experiment was repeated on the same subject and across subjects. The average reaction time across trials, denoted $s(k,l)$, was recorded. Thus $s(k,l)$ is the estimate of the (symmetrised) decision time to distinguish between $I_{k}$ and $I_{l}$. Similar estimates were obtained for several pairs of images.

\item For capturing neuronal responses to images, Sripati and Olson conducted a set of experiments on macaque monkeys. See \cite{ref:201001JNS_SriOls} for details. A single image $I_{k}$ (respectively, $I_{l}$) was displayed on the screen, and the neuronal firings elicited by $I_{k}$ (respectively, $I_{l}$) on a set of IT neurons were recorded across multiple sessions. The neuronal representation of the image $I_{k}$ was taken to be the vector of average firing rates indexed by the neurons. This is denoted $\mathbf{R}_{k} = (R_{k}(1), R_{k}(2), \ldots, R_{k}(d))$, where $d$ is the number of tapped neurons. Similarly, the neuronal representation of image $I_{l}$ was estimated and denoted as the vector $\mathbf{R}_{l}$. The measure of dissimilarity between the two images $I_{k}$ and $I_{l}$ was then taken to be the $L^{1}$-distance normalised by the number of neurons:
\begin{align}
 \label{eqn:L1 distance} \Vert \mathbf{R}_{k} - \mathbf{R}_{l}\Vert _{1} = \frac{1}{d} \sum_{m =1}^{d} \vert R_{k}(m) - R_{l} (m)\vert.
\end{align}
They obtained the scatter plot $(s(k,l)^{-1}, \Vert \mathbf{R}_{k} - \mathbf{R}_{l}\Vert _{1})_{k,l}$ shown in Fig. \ref{fig:L1_vs_Behavioural}, where $(k,l)$ varied across image pairs, and observed a remarkably high correlation $(r = 0.95)$, thereby providing evidence in support of a quantitative version of $(H)$.
\end{enumerate}

For a detailed discussion of how neural activity in monkey visual cortex can be used to predict human search performance, we refer the reader to \cite{ref:201001JNS_SriOls}, \cite{nakayama2011situating, wolfe2004attributes, palmer2000psychophysics}.

The experiments of Sripati and Olson \cite{ref:201001JNS_SriOls} and Figure \ref{fig:L1_vs_Behavioural} suggest a natural question of interest to researchers in information and decision theory. One does anticipate that $s(k,l)$ is negatively correlated with some notion of dissimilarity between $\mathbf{R}_{k}$ and $\mathbf{R}_{l}$, say $\textsf{diff}(\mathbf{R}_{k}, \mathbf{R}_{l})$.  Figure  \ref{fig:L1_vs_Behavioural} suggests
\begin{equation}
\label{qn:hypothesis 1}
 s(k,l) \cdot \Vert \mathbf{R}_{k} - \mathbf{R}_{l}\Vert _{1} = constant.
\end{equation}
However, we know of no decision theoretic basis for $\textsf{diff}(\mathbf{R}_{k}, \mathbf{R}_{l})$ to be $\Vert \mathbf{R}_{k} - \mathbf{R}_{l}\Vert _{1}$. What is an appropriate $\textsf{diff}(\mathbf{R}_{k}, \mathbf{R}_{l})$?

Familiarity with Wald's Sequential Probability Ratio Test \cite{ref:194809AMS_WalWol} immediately suggests that a relation like (\ref{qn:hypothesis 1}) should arise, but with perhaps relative entropy\footnote{This refers to relative entropy of the probability measure of a set of $d$ Poisson point processes with rate vector $\mathbf{R}_{k}$ taken with respect to the probability measure associated with rate vector $\mathbf{R}_{l}$.}, or its variant, in place of $\Vert \mathbf{R}_{k} - \mathbf{R}_{l}\Vert _{1}$. A variant may be called for because of the possibility of controlled actions. To see why, let us summarize the decision problem in the form of a question:

\begin{quotation}
  One of the six images is odd. What would the prefrontal cortex (the decision center of the brain) do if it got observations (firings of neurons) from the human analogue of the IT cortex,  and could control the eye (to gaze at one of the six objects)? The goal is to minimise the time to decide the oddball image and its location, yet keep errors within desired limits.
\end{quotation}

One can model this decision problem as a sequential hypothesis testing problem with control. Naghshvar and Javidi, earlier in \cite{ref:201006ISIT_NagJav} and more recently in \cite{ref:201312TAS_NagJav}, call such a problem active sequential hypothesis testing (ASHT). ASHT suggests a natural candidate that we shall propose for $\textsf{diff}(\mathbf{R}_{k}, \mathbf{R}_{l})$. There is however one important modeling issue that we wish to bring to the attention of the reader. Figure \ref{fig:L1_vs_Behavioural} shows that the average reaction times in the experiments are between 250 ms and 1000 ms. However, it is known that a switch in  focus from one search location to another\footnote{This rapid eye movement is called a saccade.} has a cost per switch that ranges from tens of ms to sometimes even higher than 100 ms \cite{gezeck1997saccadic}. To account for this, we extend ASHT to a setting with switching costs, and show that the $\textsf{diff}(\mathbf{R}_{k}, \mathbf{R}_{l})$ appropriate for the setting without switching costs works equally well with switching costs.

As with $L^{1}$ distance, so with our proposed $\textsf{diff}(\mathbf{R}_{k}, \mathbf{R}_{l})$, and indeed, with other natural dissimilarity indices like relative entropy and Chernoff entropy\footnote{Relative entropy and Chernoff entropy are possible candidates because of the following. Consider a simple hypothesis testing problem where exactly one of two images is displayed and the problem is to identify which. The stopping version of the problem corresponds to Wald's sequential hypothesis testing. The expected stopping time to meet a certain error tolerance criteria $\epsilon$ is roughly $\log(1/ \epsilon)/(\text{relative entropy})$ \cite{ref:194809AMS_WalWol}. When the decision is to be made after a fixed number of samples, where the number of samples is fixed upfront to meet a certain error tolerance criteria, the required number of samples is roughly $\log(1/ \epsilon)/(\text{Chernoff entropy})$.}, Table \ref{table:correlation of different neuronal metrics} indicates that all these dissimilarity measures have similar high correlation with the behavioural index\footnote{In Table \ref{table:correlation of different neuronal metrics}, correlation values are based on scatter plots arising from {\em ordered} pairs of images. This explains why $L^1$ correlation value in the table (obtained from 24 points in the scatter plot) is marginally different from the correlation indicated in Figure \ref{fig:L1_vs_Behavioural} (and obtained from 12 symmetrised points in the plot).}. Given that all these dissimilarity indices yield high correlation with the reaction times, does our proposed $\textsf{diff}$ candidate stand out in some way? It is certainly grounded in a decision-theoretic framework as we shall soon see. But is there some experimental evidence in favour of our proposed $\textsf{diff}$ candidate? We address this question as well and propose a method to rank order the dissimilarity measures in their ability to explain the experimental data of Sripati and Olson \cite{ref:201001JNS_SriOls}.

\begin{table}[t]
\caption{Correlation with Different Information Measures} % title of Table
\centering % used for centering table
\begin{tabular}{|c|c|c|c|c|c|} % centered columns (4 columns)
\hline\hline %inserts double horizontal lines
Information Measure & Correlation ($1/s(\cdots)$ vs. Discrimination index) & $p$-value \\  % inserts table
%heading
\hline % inserts single horizontal line
& & \\ % adds vertical space
Proposed & 0.94 & $5.2 \times 10^{-12}$ \\
KL & 0.93 & $8.5 \times 10^{-11}$ \\
Chernoff & 0.94 & $7.8 \times 10^{-12}$ \\
$L^1$ & 0.94 & $6.1 \times 10^{-12}$ \\ % inserting body of the table
[1ex] % [1ex] adds vertical space
\hline %inserts single line
\end{tabular}
\label{table:correlation of different neuronal metrics} % is used to refer this table in the text
\end{table}

\vspace*{.1in}

\paragraph*{Prior Work on the ASHT model}

Chernoff \cite{ref:195909AMS_Che} studied ASHT in the context of designing optimal experiments. His performance criterion was the total cost of sampling, which is proportional to delay, plus a penalty for false detection. Chernoff proposed a policy, the so-called {\it Procedure A},  and showed its asymptotic optimality as the cost of sampling went to zero. {\it Procedure A} maintains a posterior distribution on the set of hypotheses and, at each instant, selects actions according to the hypothesis with the highest posterior probability.

There has been a flurry of recent activity extending Chernoff's work in other directions. In a series of works, Naghshvar and Javidi \cite{ ref:201312TAS_NagJav, ref:201006ISIT_NagJav, ref:201010Alt_NagJav, ref:201108ISIT_NagJav, ref:201310JSP_NagJav} studied ASHT from a Bayesian cost minimization perspective. The total cost was the sum of decision delay and a penalty for false detection. They proposed policies, similar to Chernoff's {\it Procedure A}, identified bounds on the total cost, and established their proposed policies' asymptotic optimality in the same asymptotic regime as Chernoff's\footnote{They also consider the asymptotics where the number of hypotheses is large. This is not of direct relevance to our study.}. Nitinawarat et al. \cite{ref:201310ITAC_NitinVeeravalli} studied active hypothesis testing in fixed sample size and in sequential settings. They also minimize decision delay subject to a constraint on the conditional probability of false detection. When these conditional probabilities of false detection are driven to zero, the resulting asymptotic regime is the same as Chernoff's. In this asymptotic regime, they obtained results similar to those of Chernoff's but under milder assumptions. They also prove a stronger asymptotic result based on the ``risk associated with a decision''. Nitinawarat and Veeravalli \cite{2013arXiv1310.1844_NitinawaratVeeravalli} extended ASHT to Markovian observations and non-uniform costs on actions. Recently, Cohen and Zhao \cite{ref:CohenZhao_AnomalyDetection_TIT_Mar2015} studied anomaly detection from an ASHT perspective. They showed that, in their particular setting, a simple deterministic policy is asymptotically optimal. This is in contrast to random policies advocated in the other works. Further, for their particular setting, they showed the asymptotic optimality of Chernoff's policy under milder assumptions. None of the above works consider switching costs associated with a change in action.
\vspace*{.1in}

\paragraph*{Our contribution}
Broadly, our contribution is a reinterpretation of the experimental results of Sripati and Olson \cite{ref:201001JNS_SriOls} from a decision-theoretic standpoint. The following highlight some specific contributions.

\begin{itemize}
 \item We formulate the visual search problem as an ASHT problem with switching costs. We show that a modification of Chernoff's {\it Procedure A}, one that we call {\it Sluggish Procedure A}, is asymptotically optimal even with switching costs. Further, we show that the growth rate of the total cost, as the probability of false detection is driven to zero, can be made arbitrarily close to that without switching costs.
 \item We propose a neuronal dissimilarity index for the \textsf{diff} functional in lieu of the $L^{1}$ distance between the two vectors (Sripati and Olson's proposed dissimilarity index in \cite{ref:201001JNS_SriOls}). Our proposed dissimilarity index is based on, but is not the same as, the relative entropy between two Poisson point processes with the specified firing rate vectors.
 \item We test the goodness of this neuronal dissimilarity index with respect to $L^{1}$ by examining which comes closest to satisfying
     \begin{equation}
       \label{eqn:hypothesis 2}
       s(k,l) \cdot \textsf{diff}(\mathbf{R}_{k},\mathbf{R}_{l}) = constant,
     \end{equation}
     and which is farthest. We propose a comparison statistic based on the ``equality of means'' testing. We use three different equality of means tests to arrive at three different statistics. The first is the familiar ANOVA's $F$-statistic. The second is natural too, and is the analogue of the $F$-statistic associated with the family of Gamma distributed random variables instead of Gaussians. The Gamma distribution, as we will later discuss, provides a better fit for the delay data. The third is similar to the second, but assumes a known shape parameter. All three methods' rankings are consistent: our proposed dissimilarity index comes out as the best, with relative entropy coming a close second, in answer to the question: Which neural dissimilarity measure based on firing rates would be optimal from a decision-theoretic point of view? We must however add a sobering note that all three equality of means tests reject, in a rather spectacular fashion, the null hypothesis of equal means in (\ref{eqn:hypothesis 2}) at any reasonable level of statistical significance. So we emphasise that the test statistics are merely used to rank order the dissimilarity measures.
 \item Our estimation of the proposed neuronal dissimilarity index requires a near unbiased estimate of relative entropy as an intermediate step. We suggest a procedure to arrive at a nearly unbiased estimate. This maybe of independent value.
\end{itemize}

\vspace*{.1in}

\paragraph*{Organisation}

The rest of the paper is organised as follows. Section \ref{sec:ASHT problem} studies the ASHT problem with costs for switching actions. Section \ref{sec:Application to Visual Search Problem} applies the results of Section \ref{sec:ASHT problem} to the visual search problem. Section \ref{sec:ProposalForANeuronalMetric} develops the proposed neuronal dissimilarity index and discusses  its performance through correlation studies and ``equality of means'' testing. Section \ref{sec:Conclusion} provides some summarising conclusions. The proofs are relegated to appendices \ref{sec:app:PropertiesOfLLR} and \ref{proof:proof of optimum Di}. Appendix \ref{sec:app:EstimateRelEntropy} details the technique used to get a near unbiased estimate of relative entropy of one Poisson point process with respect to another.

\section{The ASHT Abstraction}
\label{sec:ASHT problem}

In this section, we describe our mathematical model for visual search and collect all the  relevant theoretical results. The development will be somewhat abstract. But we shall relate the model to visual search and shall apply the results to that setting in Section \ref{sec:Application to Visual Search Problem}. The main contribution of this section is the asymptotic growth rate of cost. In Section \ref{sec:ProposalForANeuronalMetric}, we shall see how this suggests an appropriate \textsf{diff} function for plugging into (\ref{eqn:hypothesis 2}).

\subsection{The ASHT Model}

\subsubsection{The model description}
\label{sec:basic notation}

Let us begin by setting up some notation.

Let $H_{i}$, $i = 1,2, \ldots,M$ denote the $M$ hypotheses of which exactly one, denoted $H$, holds true. In this section, we do not assume a prior on the hypotheses. Let $\mathcal{A}$ be the set of all possible actions which we take as finite: $|\mathcal{A}| = K < \infty$. Let $\mathcal{X}$ be the observation space. Let $(X_{n})_{n \ge 1}$ and $(A_{n})_{n \ge 1}$ denote the observation process and the control process respectively.  We write $X^n$ for $(X_1, \ldots, X_n)$ and similarly $A^n$ for $(A_1, \ldots, A_n)$. We also write $\mathcal{P}(\mathcal{A})$ for the set of probability distributions on $\mathcal{A}$.

A policy $\pi$ is a sequence of action plans that at time $n$ looks at the history ${X}^{n-1},{A}^{n-1}$ and prescribes a composite action that is either $(stop, \delta)$ or $(continue, \lambda)$ as explained next. If the composite action is $(stop, \delta)$, then the controller stops taking further samples (or retires) and indicates $\delta$ as its decision on the hypothesis; $\delta \in \{1, 2, \ldots, M\}$. If the composite action is $(continue, \lambda)$, the controller picks the next action $A_{n}$ according to the distribution $\lambda \in \mathcal{P}(\mathcal{A})$. Let $\tau(\pi)$ be the stopping time $$\tau(\pi) := \inf\{n \geq 1 | A_{n} = (stop,\cdot)\}.$$

Consider a policy $\pi$. Conditioned on action $A_{n}$ and  the true hypothesis $H$, we assume that $X_{n}$ is conditionally independent of previous actions ${{A}^{n-1}}=(A_{1}, A_{2},\dots, A_{n-1})$, previous observations  ${{X}^{n-1}}=(X_{1}, X_{2}, \dots, X_{n-1})$, and the policy. Let $q_{i}^{a}$ be the conditional probability density function, with respect to some reference measure $\mu$, of the observation $X_{n}$ under action $a$ when $H = H_{i}$. Let $D(q_{i}^{a}\Vert q_{j}^{a})$ denote the relative entropy\footnote{By an abuse of notation, we use the densities of the probability measures as the arguments of the relative entropy function.} between the conditional probability measures associated with the observations under hypothesis $H_{i}$ and under hypothesis $H_{j}$, upon action $a$. Denote by  \textsf{unif($\mathcal{A}$)} the uniform distribution on $\mathcal{A}$. Let $q_{i}^{\pi}({x^{n}},{a^{n}})$ be the probability density function of observations and actions $({x^{n}},{a^{n}})$ till time $n$ under policy $\pi$, with respect to the common reference measure $\mu^{\otimes n} \times \textsf{unif($\mathcal{A}$)}^{\otimes n}$. Let $Z_{i}^{\pi}(n)$ denote the log-likelihood process of hypothesis $H_{i}$, i.e.,
\begin{align}
 \label{eqn:general LLP}   Z_{i}^{\pi}(n) &= \log {q_{i}^{\pi}\left({{X}^{n}},{{A}^{n}} \right)}.
\end{align}
 Going forward, for ease of notation, we drop the superscript $\pi$ while describing $q_{i}^{\pi}$, $Z_{i}^{\pi}$, and other variables, but their dependence on the underlying policy should be kept in mind, and the policy under consideration will be clear from the context.
Define $Z(n) = (Z_{1}(n), Z_{2}(n), \ldots, Z_{M}(n))$.
Let $Z_{ij}(n)$ denote the log-likelihood ratio (LLR) process of $H_{i}$ with respect to $H_{j}$, i.e.,
\begin{align*}
  Z_{ij}(n) &= Z_{i}(n) - Z_{j}(n)\\
  \nonumber &= \log \frac{q_{i}\left({{X}^{n}},{{A}^{n}} \right)}{q_{j}\left({{X}^{n}},{{A}^{n}} \right)}\\
  \nonumber &=  \sum_{l=1}^{n}\log\frac{ q_{i}^{A_{l}}\left(X_{l} \right)}{  q_{j}^{A_{l}}\left(X_{l} \right)}.
\end{align*}
Let $E_{i}$ denote the conditional expectation and let $P_{i}$ denote the conditional probability measure under $H = H_{i}$. (More formally, these should be represented $E_{i}^{\pi}$ and $P_{i}^{\pi}$. But as done above, we omit the superscript $\pi$.)

Given an error tolerance vector $\alpha = (\alpha_{1}, \alpha_{2}, \ldots, \alpha_{M})$ with $0<\alpha_{i}<1$, let $\Pi(\alpha)$ be the set of policies
\begin{align*}
 \Pi(\alpha)= \left\{\pi: P_{i}(\delta \ne i) \le \alpha_{i}, \; \forall \; i \right\}.
\end{align*}
These are policies that meet a specified tolerance for the conditional probability of false detection. We define $\Vert \alpha \Vert := \max_i \alpha_i$.

We define $\lambda_{i}$ to be the best mixed action that guards $H_{i}$ against its nearest alternative\footnote{ This suffices because the probability of error is dominated by the nearest alternative hypothesis.}, i.e., $\lambda_{i} \in \mathcal{P}(\mathcal{A})$ such that
  \begin{align}
   \label{eqn:optimal_lambda}
   \lambda_{i} := \arg \max_{\lambda \in \mathcal{P}(\mathcal{A})} \left[ \min_{j \ne i} \sum_{a  \in \mathcal{A}} \lambda(a) D(q_{i}^{a}\Vert q_{j}^{a})\right].
  \end{align}
If there are several maximizers, pick one arbitrarily. Further, define
 \begin{align}
\label{eqn:D_i}
  D_{i} := \max_{\lambda \in \mathcal{P}(\mathcal{A})} \left[\min_{j \ne i} \sum_{a \in \mathcal{A}} \lambda({a}) D\left(q_{i}^{a}\Vert q_{j}^{a}\right)\right].
\end{align}
Let $\mathcal{A}_{ij} := \{a \in \mathcal{A}: D(q_{i}^{a}\Vert q_{j}^{a})>0\}$, the set of all actions that can differentiate hypothesis $H_{i}$ from hypothesis $H_{j}$. Since $D(q_{i}^{a}\Vert q_{j}^{a})=0 \Leftrightarrow  D(q_{j}^{a}\Vert q_{i}^{a})=0$, we have $\mathcal{A}_{ij} = \mathcal{A}_{ji}$.

\subsubsection{Assumptions}

Throughout, we make the following assumptions.
\begin{enumerate}
%  \item $D(q_{i}^{a}\Vert q_{j}^{a}) < \infty \qquad \forall i,j,a$.
% \item For every $i,j, \text{ such that } j \ne i$, there is a control $a \in \mathcal{A}$ such that $D(q_{i}^{a}\Vert q_{j}^{a}) > 0$.
\item [(I)\;\;  ] $E_{i}\left[\left(\log \frac{q_{i}^{a}(X)}{q_{j}^{a}(X)} \right)^{2}\right] < \infty$ $\forall$ $i,j,a$.\\
\item [(IIa)] \label{assumpIIa} $\mathcal{A}_{ij} \ne \varnothing \quad \forall i,j \mbox{ such that } i \ne j$, and
\item [(IIb)] \label{assumpIIb} $\beta := \min \left\{ \sum_{a \in \mathcal{A}_{ij}} \lambda_{k}(a) ~|~ 1 \leq i, j, k \leq M, ~ i \ne j \right\} > 0$.
\end{enumerate}

Assumption (I) implies that $D(q^a_i || q^a_j) < \infty$, which in turn ensures that no single observation can result in a reliable decision. Assumption (I) is used in proving the lower bound on the expected number of samples needed to satisfy the tolerance criterion. This is also assumed by Chernoff \cite{ref:195909AMS_Che} and Nitinawarat et al. \cite{ref:201310ITAC_NitinVeeravalli}.

Assumption (IIa) ensures that for any distinct $i$ and $j$, there is at least one control that can help distinguish the hypotheses $H_{i}$ from $H_{j}$. If $\mathcal{A}_{ij} = \emptyset$ for some $i$ and $j$, it is impossible to distinguish them from each other. Assumption (IIb) is a stronger assumption than, and implies, Assumption (IIa). Assumption (IIb) ensures that if actions are taken according to any of the $\lambda_{k}$ in (\ref{eqn:optimal_lambda}) then, for any pair of hypotheses $H_{i}$ and $H_{j}$, there is a positive probability of choosing an action that can discriminate the pair. We shall use Assumption (IIb) in the achievability proofs of our policies. It allows for easier proofs for our policies, and makes the presentation simpler. However one can work with Assumption (IIa) as well, and construct asymptotically optimal policies, with minor modifications to our policies. We will describe the needed modifications later in this section.

%-----------------------------------%
\subsubsection{Switching cost and total cost}

The costs are as follows.

\paragraph*{Switching Cost}
Let $g(a,a')$ denote the cost of switching from action $a$ to action $a'$. Throughout, we make the following additional assumptions.
\begin{itemize}
  \item[(III)] $g(a,a') \ge 0 \quad \forall a,a' \in \mathcal{A}$, $g({a,a}) = 0 ~ \forall a \in \mathcal{A}$, and $g_{\max} := \max_{a,a'} g(a,a') < \infty$.
\end{itemize}
The assumption in the middle says no switching incurs zero cost. This assumption will play a crucial role towards our eventual conclusion that switching costs do not matter in the asymptotics considered in this paper.

\paragraph*{Total cost}
\label{sec:performance criterion}
For a policy $\pi \in \Pi(\alpha)$, the total cost $C(\pi)$ is taken to be the sum of the stopping time (delay) and the net switching cost, i.e.,  $$C(\pi) := \tau(\pi)+ \sum_{l=1}^{\tau(\pi)-1} g({A_{l},A_{l+1}}).$$

\subsubsection{Asymptotics}
We shall be interested in the asymptotics of the minimum expected total cost $E_{i}[C(\pi)]$, minimized over policies in $\Pi(\alpha)$, as $||\alpha|| \rightarrow 0$. Note that there are $M$ such conditional expected total costs, one for each hypothesis.

\subsection{Results on the ASHT Model}
We collect all the main results in this section. We first identify a lower bound.
\subsubsection{The converse - Lower bound}
\label{sec:lower bounds}
The following proposition gives a lower bound on the conditional expectation of the stopping time, given hypothesis $H = H_{i}$, for all policies belonging to $\Pi(\alpha)$.
\begin{proposition}
 \label{proposition:proposition1}
Assume (I). For each $i$, we have
 \begin{align}
\label{eqn:lower bound} \lim _{\Vert \alpha \Vert \rightarrow 0} \inf_{\pi \in \Pi(\alpha)} \frac{E_{i} [\tau(\pi)]}{\vert \log \Vert \alpha \Vert \vert} \ge \frac{1}{D_{i}},
 \end{align}
where $D_{i}$ is given in (\ref{eqn:D_i}).
 \end{proposition}
\begin{IEEEproof}
 Since only expected time to stop is considered, proof of \cite[Th. 2, p. 766]{ref:195909AMS_Che} applies.
%(Alternatively, see proof of \cite[Lem. 2.1, Th. 2.2]{ref:1998SISP_Tar}).
\end{IEEEproof}
\vspace*{0.1 in}
We then have the following corollary.
\begin{corollary}
Assume (I). For each $i$, we have
{\small
\label{cor:total_cost_lower_bound}
 \begin{align}
\label{eqn:total_cost_lower_bound} \lim_{\Vert \alpha \Vert \rightarrow 0} \inf_{\pi \in \Pi(\alpha)} \frac{E_{i}[ C(\pi)]}{|\log \Vert \alpha \Vert |} \ge \frac{1}{D_{i}}.
 \end{align}
 }
\end{corollary}

\begin{IEEEproof}
 With switching costs added, we have $C(\pi) \ge \tau(\pi)$, and the corollary follows from Proposition \ref{proposition:proposition1}.
\end{IEEEproof}

\vspace*{0.1 in}

\subsubsection{Achievability -  A modification to Chernoff's Procedure A}
\label{sec:upper bounds}
Chernoff \cite{ref:195909AMS_Che} proposed a policy termed {\it Procedure A} and showed that it has asymptotically optimal expected decision delay. We now describe {\it Procedure A}.

\begin{addmargin}[2em]{2em}% 1em left, 2em right
% \begin{figure}[ht]
% \quad\\
{\it Policy {\it Procedure A:}} $\pi_{PA}(L)$ \\Fix $L > 0 $. \\
At time $n$:
\begin{itemize}
\item Let $\theta(n) = \arg\max_{i} Z_{i}(n)$, the index with the largest log-likelihood at the current time. Ties are resolved uniformly at random.
\item If $ Z_{\theta(n),j}(n) < \log{((M-1)L)}$ for some $j \ne \theta(n)$ then $A_{n+1}$ is chosen according to $\lambda_{\theta(n)}$, i.e.,
\begin{align}
\label{eqn:probability of action selection}
 \Pr(A_{n+1} = a) = \lambda_{\theta(n)}(a).
\end{align}
\item If $Z_{\theta(n),j}(n) \ge \log{((M-1)L)}$ for all $j \ne \theta(n)$ then the test retires and declares $H_{\theta(n)}$ as the true hypothesis.
\end{itemize}
% \end{figure}
\end{addmargin}
\vspace*{0.1in}

We now describe a modified policy that comes arbitrarily close to being asymptotically optimal in the presence of switching costs. We introduce a switching parameter $\eta, \: 0 < \eta \le 1 $, which determines the maximum transition rate out of a given action. When $\eta = 1$, we will have the original {\it Procedure A}. When $\eta$ approaches zero, the rate of jumping out of the current action approaches zero.
\begin{addmargin}[2em]{2em}
{\it Policy {\it Sluggish Procedure A:}} $\pi_{SA}(L,\eta)$ \\Fix $L > 0 , \: 0 < \eta \le 1$. \\
At time $n$:
\begin{itemize}
\item Let $\theta (n) = \arg\max_{i} Z_{i}(n)$. Ties are resolved uniformly at random.
\item If $ Z_{\theta(n),j}(n) < \log({(M-1)L})$ for some $j \ne \theta(n)$ then $A_{n+1}$ is chosen as follows.
\begin{itemize}
 \item Generate $U_{n+1}$, a Bernoulli($\eta$) random variable, independent of all other random variables.
 \item If $U_{n+1} = 0$, then $A_{n+1} = A_{n}$.
 \item If $U_{n+1} = 1$, then generate $A_{n+1}$ according to distribution $\lambda_{\theta(n)}$.
\end{itemize}

\item If $Z_{\theta(n),j}(n) \ge \log{(M-1)L}$, for all $j \ne \theta(n)$, then the test retires and declares $H_{\theta(n)}$ as the true hypothesis.
\end{itemize}
\end{addmargin}
\vspace*{0.1in}

We also consider two variants of $\pi_{SA}(L,\eta)$ which are useful in the analysis.
\begin{itemize}
 \item {\it Policy} $\pi_{SA}^{i}(L,\eta)$: This is the same as $\pi_{SA}(L,\eta)$, but stops only at decision $i$ when $\min_{j:j \ne i} Z_{ij}(n) \ge \log(L (M-1))$.
 \item {\it Policy} $\tilde{\pi}_{SA}(\eta)$: This is the same as $\pi_{SA}(L,\eta)$, but never stops, and hence $L$ is irrelevant.
\end{itemize}
Under a fixed hypothesis $H = H_{i}$, and the triplet of policies $(\pi_{SA}(L,\eta), \pi_{SA}^{i}(L,\eta), \tilde{\pi}_{SA}(\eta))$, it is easily seen that there is a common underlying probability measure with respect to which the processes $(X_n, A_n)_{n \ge 1}$ associated with the three policies are naturally coupled, with only the stopping times being different. Under this coupling, the following are true:
\begin{align*}
 \tau(\pi_{SA}^{i}(L,\eta)) & \ge \tau(\pi_{SA}(L,\eta)),\\
 \{\tau (\pi_{SA}(L,\eta)) > n\} &\subset \{\tau(\pi_{SA}^{i}(L,\eta)) > n\}\\
&\subset \left\{\min_{j:j \ne i} Z_{ij}(n) < \log(L (M-1))\right\}.
 \end{align*}

Policy $\pi_{SA}(L,\eta)$ is designed to stop only when the posteriors suggest a reliable decision. This is formalized now.
\begin{proposition}
 \label{prop:probability of wrong detection}
Assume (I) and (IIb). For Policy ${\pi_{SA}(L,\eta)}$, the conditional probability of error under hypothesis $H_{i}$ is upper bounded by $P_{i}(\delta \ne i) \le 1/L.$
% \begin{align}
%  P_{i}(\delta \ne i) \le \frac{1}{L}.
% \end{align}
\end{proposition}

See Appendix \ref{appendix:proof of probability of wring detection} for a proof. As a consequence we have ${\pi_{SA}}(L,\eta) \in \Pi(\alpha)$ if $\alpha_{i} \ge 1/L$ for every $i$.

We now state the time-delay performance of the policy $\pi_{SA}(L,\eta)$.

\begin{thm}
 Assume (I) and (IIb). Consider the policy $\pi_{SA}(L, \eta)$. The conditional expected time to make a decision, for each $i$, satisfies
\label{theorem: upper bound on stopping time of SA}
\begin{align}
\label{eqn:upper_bound_sandwich_argument}
\lim_{L \rightarrow \infty} \frac{E_{i}\left[\tau(\pi_{SA}(L,\eta))\right]}{\log L} \le \frac{1}{D_{i}}.
\end{align}
\end{thm}

See Appendix \ref{ref:lemma:achievability proof} for a detailed proof. This result will be crucial because the policy $\pi_{SA}(L, \eta)$, despite its sluggishness induced by $\eta$, remains asymptotically optimal when only the stopping time $\tau(\pi_{SA}(L,\eta))$ is considered as cost. We now leverage this to show that, if $\eta$ is sufficiently small, $\pi_{SA}(L,\eta)$ is near optimal when switching costs are also taken into account.

\begin{proposition}
 \label{prop:achievability}
 Assume (I), (IIb), and (III). Consider the policy $\pi_{SA}(L, \eta)$. We then have, for each $i$,
 \begin{align}
 \label{eqn:achievability}
  \lim _{L \rightarrow \infty} E_{i}\left[\frac{C(\pi_{SA}(L,\eta))}{\log L}\right] & \le \frac{1}{D_{i}} + \frac{g_{\max}\eta }{D_{i}}.
 \end{align}
\end{proposition}

\begin{IEEEproof}
We can write the following chain of inequalities.
\begin{align}
\nonumber E_{i} &\left[C(\pi_{SA}(L,\eta))\right]\\
\nonumber  &= E_{i}\left[\tau(\pi_{SA}(L,\eta))+\sum_{l=1}^{\tau(\pi_{SA}(L,\eta))-1} g(A_{l},A_{l+1})\right]\\
\nonumber & \le E_{i}\left[\tau(\pi_{SA}(L,\eta))\right]+g_{\max}E_{i}\left[\sum_{l=1}^{\tau(\pi_{SA}(L,\eta))-1}1_{\{A_{l}\ne A_{l+1}\}}\right]\\
\nonumber & \le E_{i}\left[\tau(\pi_{SA}(L,\eta))\right]+g_{\max}E_{i}\left[\sum_{l=1}^{\tau(\pi_{SA}(L,\eta))-1}U_{l+1}\right]\\
\nonumber & = E_{i}\left[\tau(\pi_{SA}(L,\eta))\right]+g_{\max}\eta E_{i}\left[\tau(\pi_{SA}(L,\eta))-1\right]\\
\label{eqn:cost_bound}& \le E_{i}\left[\tau(\pi_{SA}(L,\eta))\right](1+g_{\max}\eta).
\end{align}
In the above chain, the second inequality follows from Assumption (III). The penultimate equality holds because of Wald's equation \cite{ref:wald1944}. Dividing by $\log L$, letting $L \rightarrow \infty$, and using Theorem \ref{theorem: upper bound on stopping time of SA}, we see that (\ref{eqn:achievability}) holds.
\end{IEEEproof}

\subsubsection{Asymptotic optimality}

Corollary \ref{cor:total_cost_lower_bound} and Proposition \ref{prop:achievability} show that, when the conditional probability of false detection is driven to zero, the proposed policy $\pi_{SA}(L,\eta)$ has nearly the same growth rate for cost as an asymptotically optimal policy without switching costs. We now make the above statement precise.  The parameter $\eta$ should be suitably chosen to get sufficiently close to asymptotic optimality.

\begin{thm}
\label{thm:asymptotic optimality of sluggish procedure A}
 Assume (I), (IIb), and (III). Consider a sequence of vectors $(\alpha^{(n)})_{n \ge 1}$, where  $\alpha^{(n)}$ is the $n^{th}$ tolerance vector, such that  $\lim_{n \rightarrow \infty} \Vert \alpha^{(n)} \Vert = 0$ and
 \begin{align}
 \label{eqn:condition on max and min of tolerance vector}
   \lim_{n \rightarrow \infty} \frac{\Vert \alpha^{(n)} \Vert}{\min_{k}\alpha^{(n)}_{k}} < B
 \end{align} for some $B$. Then, for each $n$, the policy $\pi_{SA}(L_n,\eta)$ with $\log L_n = -\log \min_{k}\alpha^{(n)}_{k}$ belongs to $\Pi(\alpha^{(n)})$. Furthermore, for each $i$,
\begin{align}
\label{eqn:asymptotic optimality}
 \lim_{n \uparrow \infty}\inf_{\pi \in \Pi(\alpha^{(n)})} \frac{E_{i}\left[C(\pi)\right]}{\log L_n} = \lim_{\eta \downarrow 0}\lim_{n \uparrow \infty} \frac{E_{i}\left[C(\pi_{SA}(L_n,\eta))\right]}{\log L_n} = \frac{1}{D_{i}}.
\end{align}
\end{thm}

\begin{IEEEproof}
 The fact that $\pi_{SA}(L_n,\eta) \in \Pi(\alpha^{(n)})$ is evident from Proposition \ref{prop:probability of wrong detection}, and $1 / L_{n} \le \alpha^{(n)}_{k}, \: k = 1,2, \cdots, n$. We then have the following chain of inequalities:
\begin{align*}
 \frac{1}{D_{i}} &\le  \lim_{n \uparrow \infty}\inf_{\pi \in \Pi(\alpha^{(n)})} \frac{E_{i}\left[C(\pi)\right]}{|\log \Vert \alpha^{(n)} \Vert|}\\
 & = \lim_{n \uparrow \infty}\inf_{\pi \in \Pi(\alpha^{(n)})} \frac{E_{i}\left[C(\pi)\right]}{\log L_n}\\
 & \le \lim_{\eta \downarrow 0}\lim_{n \uparrow \infty} \frac{E_{i}\left[C(\pi_{SA}(L_n,\eta))\right]}{\log L_n}\\
 & \le \frac{1}{D_{i}}.
\end{align*}
The first inequality follows from Corollary \ref{cor:total_cost_lower_bound}. The next equality follows from the fact that
\[
  \lim_{n \rightarrow \infty}\frac{|\log \Vert \alpha^{(n)} \Vert |}{\log L_{n}} = 1,
\]
which in turn is true due to the assumption (\ref{eqn:condition on max and min of tolerance vector}). The third inequality follows because $\pi_{SA}(L_{n},\eta)$ is one specific policy in $\Pi(\alpha^{n})$. The last inequality follows from Proposition \ref{prop:achievability} after letting $\eta \downarrow 0$. Consequently, all inequalities must be equalities.
\end{IEEEproof}

%%%%%%%%%%%%%%%%%%%%%%%%%%%%%%%%%%%%%%%%%%%%%%%%%%%%%%%%%%%%%%%%%%%%%%%%%%%%%%%%%%%%%%%%%%%%
\subsection{Discussion on Assumption (IIb)}
\label{sec:Assumption (II) discussion}

Chernoff's result on the asymptotic optimality of {\it Procedure A} \cite{ref:195909AMS_Che} was proved under a stronger assumption than Assumption (IIb), namely, Chernoff required
\begin{align}
\label{eqn:Chernoff's assumption}
D(q_{i}^{a} \Vert q_{j}^{a}) > 0  \text{ for all } a \text{ and for all pairs $i \ne j$}.
\end{align}
Assumption (IIb) ensures that, at all times, and for any pair of hypotheses $i$ and $j$, $i \ne j$, there is a positive probability of choosing an action that can distinguish the two hypotheses. This suffices for Chernoff's proofs to go through. Specifically, we shall use Assumption (IIb) to prove the exponential decay result in Proposition \ref{prop:exponential decay LLR process} of Appendix \ref{sec:app:PropertiesOfLLR}. Nitinawarat et al. \cite{ref:201310ITAC_NitinVeeravalli} proposed a modified {\it Procedure A} that sampled actions randomly at intervals $\lceil{{\nu}^{l}\rceil}_{l \ge 1}, \; \nu > 1$, and showed that their proposed policy is asymptotically optimal under the weaker Assumption (IIa). The random sampling enabled them to obtain a polynomial decay counterpart of Proposition \ref{prop:exponential decay LLR process} of Appendix \ref{sec:app:PropertiesOfLLR}. Recently, Cohen and Zhao \cite{ref:CohenZhao_AnomalyDetection_TIT_Mar2015} showed the asymptotic optimality of {\it Procedure A} under the weaker Assumption (IIa) for an active anomaly detection problem, which is a specific ASHT problem. We conjecture that Chernoff's {\it Procedure A} is asymptotically optimal under the weaker Assumption (IIa) for all ASHT problems. A proof of this claim has remained elusive. Nevertheless, policies whose performances are provably arbitrarily close to the optimum can be designed. We make the above claim precise in the next proposition.

\begin{proposition}
 \label{prop: Near optimal provable policies}
 Assume (I) and (IIa). Fix $\epsilon > 0$. Then there exists a sequence of policies $\{\pi_{\epsilon}(L)\}$ that satisfies $\pi_{\epsilon}(L) \in \Pi(\frac{1}{L},\frac{1}{L},\cdots,\frac{1}{L})$ and
 \begin{align}
  \lim _{L \rightarrow \infty} E_{i}\left[\frac{\tau(\pi_{\epsilon}(L))}{\log L}\right] & \le \frac{1}{(1-\epsilon)D_{i}}.
 \end{align}
\end{proposition}
We omit the proof because the needed modifications to the proof of Theorem \ref{theorem: upper bound on stopping time of SA} are straightforward. Policy $\{\pi_{\epsilon}(L)\}$ can be constructed as a variant of {\it Procedure A} that, at each instant $n$, chooses an action according to $\textsf{unif}(\mathcal{A})$ with probability $\epsilon$ or according to (\ref{eqn:probability of action selection}) with probability $(1- \epsilon)$.
Note that, at each time $n$, the modified policy $\{\pi_{\epsilon}(L)\}$ uses a randomisation on the actions of the form $\tilde{\lambda}_{\theta(n)} = (1-\epsilon)\lambda_{\theta(n)}+\epsilon \textsf{unif($\mathcal{A}$)}$. It can be shown that, under hypothesis $H_{i}$, $\theta(n) = i$ in finite time with probability 1, and thereby the asymptotic log likehood ratio rate between $H_{i}$ and any other $H_{j}$ will be lower bounded by $(1- \epsilon) D_{i}$. Thus, at the cost of a small penalty, we can design nearly asymptotically optimal policies under the weaker Assumption (IIa). A similar argument holds true with switching costs, just as Theorem \ref{theorem: upper bound on stopping time of SA} is extended in Theorem \ref{thm:asymptotic optimality of sluggish procedure A}, albeit with a corresponding but arbitrarily small increase in the total cost. Again, we omit the proof of this claim with switching costs. The conclusion is that Assumption (IIa) suffices for the asymptotic growth rate to be $\frac{1}{D_{i}}$.
%%%%%%%%%%%%%%%%%%%%%%%%%%%%%%%%%%%%%%%%%%%%%%%%%%%%%%%%%%%%%%%%%%%%%%%%%%%%%%%%%%%%%%%%%%%%

\section{Back to Visual Search}
\label{sec:Application to Visual Search Problem}

We now return to the visual search problem. In the visual search task, a subject has to identify an oddball image from amongst $W$ images displayed on a screen  ($W = 6$ in Figures \ref{fig:fig1a} - \ref{fig:fig1b}). For the purpose of modeling, we make the following assumptions. The subject can focus attention on only one of the $W$ positions, and the field of view is restricted to the image at that position alone. Further, we assume that time is slotted\footnote{One could also consider an extension to the continuous-time setting. But all essential ideas are best described in the slotted setting.} and each slot is of duration $T$. The subject can change the focus of his attention to any of the $W$ image locations, but only at the slot boundaries. {A switch in focus of attention (saccade) requires an integer number of slots for the operation, and no sensing is possible during such a saccade. The lost time during saccades are modeled as switching costs (delays), and hence the total  decision time  is the sum of sensing delay and switching delays.} We assume that the subject would have indeed found the exact location and identity of the oddball image before mapping it to a ``left'' or ``right'' decision. These are clearly oversimplifying assumptions, but enable easier analysis and provide valuable insights.

If the image in the focused location is $I_{k}$, we assume that a set of $d$ neurons react accordingly to produce spike trains. These constitute the observations. Specifically, these are modeled as $d$ independent Poisson point processes of duration $T$ with rates given by the components of the rate vector $\mathbf{R}_{k} = (R_{k}(1), R_{k}(2), \ldots, R_{k}(d))$. More formally, let $\mathcal{X}$ be the space of counting processes in $[0,T]$ with an associated $\sigma$-algebra. Let $\mu_{1,T}$ be the standard rate $1$ Poisson point process and let $\mu_{1,T}^{\otimes d}$ be its $d$-fold product measure. Let $\mu_{\mathbf{R}_{k},T}$ denote the probability measure $P_{k}$, so that density of $\mu_{\mathbf{R}_{k},T}$ with respect to $\mu_{1,T}^{\otimes d}$ is given by
\[
  f_{k} := \frac{d\mu_{\mathbf{R}_{k},T}}{d\mu_{1,T}^{\otimes d}},
\]
with a similar definition for $f_{l}$ corresponding to image $I_{l}$.

We now describe two possible settings. Case 2 will turn out to be closer to the experiment of Sripati and Olson \cite{ref:201001JNS_SriOls}.
\vspace*{0.1 in}

{\em Case 1: The subject has knowledge that the oddball image is $I_{k}$ and that the distractors are $I_{l}$. Since there are $W$ locations, and $1\le i \le W$, there are $W$ hypotheses.}
\vspace*{0.1 in}

\label{sec:visual search known odd and dist dists}
The visual search problem under Case 1 can be formulated as an ASHT problem as follows.

\begin{itemize}
 \item Hypotheses: $H_{i}$ is the hypothesis that the oddball image $(I_{k})$ is at location $i$, $1 \le i \le W$.
 \item Actions: The subject may focus on any one of the $W$ locations, and so $\mathcal{A} = \{1,2, \ldots, W\}$.
 \item Observations: The conditional probability density function $q_{i}^{a}$ of the observations, under hypothesis $H_{i}$ and when action $a$ is chosen, is:
\begin{align*}
%  \label{eqn:kernels1}
 q_{i}^{a} &=
\begin{cases}
f_{k} & \quad \text{if $a = i$}\\
f_{l} & \quad \text{if $a \ne i$.}
\end{cases}
\end{align*}
\end{itemize}

In words, under Hypothesis $H_i$, the oddball image is $I_{k}$ and is at location $i$. If the action is to focus on location $i$, i.e., $a = i$, then the subject views the oddball image $I_{k}$, and so the observations have density $f_{k}$. If $a \ne i$, then the subject views the distractor image $I_{l}$, and so the observations have density $f_{l}$.

The relative entropies for the various combinations of hypotheses pairs $(i,j)$, with $i \ne j$, and actions are as follows:

\begin{align}
\label{eqn: Case1 KL distances between hypotheses under action a}
 &D(q_{i}^{a} \Vert q_{j}^{a}) =
\begin{cases}
 D(f_{k} \Vert f_{l})& \text{$a=i$}\\
 D(f_{l} \Vert f_{k})& \text{$a=j$}\\
 0& \text{$a\ne i$, $a \ne j$}.
\end{cases}
\end{align}

\begin{proposition}
\label{prop:optimum Di Case1}
 For the setting of Case 1, the $\lambda_{i}$ and $D_{i}$ of (\ref{eqn:optimal_lambda}) and (\ref{eqn:D_i}), respectively, are as follows.\\
 If $D(f_{k}\Vert f_{l}) > D(f_{l}\Vert f_{k}) / (W-1)$ then
\begin{align*}
 \nonumber \lambda_{i}(i)&=1, \lambda_{i}(j) = 0 \: \forall j \ne i, ~and ~ D_{i} = D(f_{k}\Vert f_{l}).
\end{align*}
If $D(f_{k}\Vert f_{l}) \le D(f_{l}\Vert f_{k}) / (W-1)$ then
\begin{align*}
\nonumber \lambda_{i}(i)&=0,
\nonumber \lambda_{i}(j)=\frac{1}{(W-1)} \: \forall j \ne i, ~ and ~ D_{i} = \frac{D(f_{l}\Vert f_{k})}{(W-1)}. %\label{eqn: optimum lambda 4}
\end{align*}
\end{proposition}

\begin{IEEEproof}
We can upper bound $D_{i}$ as follows:
 \begin{align}
  \label{eqn:Case1 lambda_i 1}D_{i} &= \max_{\lambda \in \mathcal{P}(\mathcal{A})} \min_{j \ne i} \sum_{a \in \mathcal{A}} \lambda(a) D(q_{i}^{a} \Vert q_{j}^{a})\\
  \label{eqn:Case1 lambda_i 2}&=\max_{\lambda \in \mathcal{P}(\mathcal{A})} \min_{j \ne i}  \left[\lambda(i) D(f_{k} \Vert f_{l})+\lambda(j) D(f_{l} \Vert f_{k})\right]\\
  \label{eqn:Case1 lambda_i 3}&=\max_{\lambda \in \mathcal{P}(\mathcal{A})} \left[\lambda(i) D(f_{k} \Vert f_{l}) +\min_{j \ne i} \lambda(j) D(f_{l} \Vert f_{k})\right]\\
  \label{eqn:Case1 lambda_i 4}&\le \max_{\lambda \in \mathcal{P}(\mathcal{A})} \left[\lambda(i) D(f_{k} \Vert f_{l}) +\frac{1 - \lambda(i)}{W-1} D(f_{l} \Vert f_{k})\right]\\
  \label{eqn:Case1 lambda_i 5}&= \begin{cases}
 D(f_{k} \Vert f_{l})& \text{if } D(f_{k} \Vert f_{l}) \ge \frac{D(f_{l} \Vert f_{k})}{W-1}, \text{by setting $\lambda(i) = 1$,}\\
 \frac{D(f_{l} \Vert f_{k})}{W-1}& \text{if } D(f_{k} \Vert f_{l}) < \frac{D(f_{l} \Vert f_{k})}{W-1},\text{by setting $\lambda(i)=0$}.
\end{cases}
 \end{align}
 Here (\ref{eqn:Case1 lambda_i 2}) follows from (\ref{eqn: Case1 KL distances between hypotheses under action a}), (\ref{eqn:Case1 lambda_i 3}) follows after taking the minimisation inside, (\ref{eqn:Case1 lambda_i 4}) follows because the minimum of a set of numbers is upper bounded by their arithmetic mean, and (\ref{eqn:Case1 lambda_i 5}) follows by maximising the linear objective function in (\ref{eqn:Case1 lambda_i 4}). Finally, (\ref{eqn:Case1 lambda_i 4}) can be made an equality by choosing all $\lambda_{j}, j \ne i$ to be identical. This proves the Proposition.
\end{IEEEproof}

Thus, under $H_{i}$, to distinguish $H_{i}$ from its nearest alternative, one either focuses only at the oddball location or at any of the other locations with equal probability depending on whether $D(f_{k}\Vert f_{l}) > D(f_{l}\Vert f_{k}) / (W-1)$ or not.

\vspace*{0.1 in}

{\em Case 2: The subject has knowledge of the two competing images $I_{k}$ and $I_{l}$, but does not know which of the two is the oddball image.}

\label{sec:visual search with knowledge of odd and dist dists}
\label{para_case2}

This visual search problem can be formulated as a $2W$ hypothesis testing problem as follows.
\begin{itemize}
 \item Hypotheses:
\begin{align*}
H_{i} \text{ with } i \le W &: \text{ The oddball image is $I_{k}$ and is at location $i$. All other locations have image $I_{l}$}.\\
H_{i} \text{ with } i > W &: \text{ The oddball image is $I_{l}$ and is at location $i-W$. All other locations have image $I_{k}$}.
\end{align*}

\item Actions: The subject can focus on any one of the $W$ locations, and so $\mathcal{A} = \{1, 2, \cdots, W\}$.

\item Observations:  The conditional probability density function $q_{i}^{a}$ of the observations, under hypothesis $H_{i}$ and when action $a$ is chosen, is:

\begin{align*}
%  \label{eqn:kernels}
 q_{i}^{a} &=
\begin{cases}
f_{k} & \text{$i \le W$, $a = i$}\\
f_{l} & \text{$i \le W$, $a \ne i$}
\end{cases}\\
 q_{i}^{a} &=
\begin{cases}
f_{l} & \text{$i > W$, $a = i - W $}\\
f_{k} & \text{$i > W$, $a \ne i - W$}.
\end{cases}
\end{align*}
\end{itemize}

In words, under Hypothesis $H_i$ with $i \leq W$, the oddball image is $I_k$ and is at location $i$. If the action is to focus on location $i$, i.e., $a = i$, then the subject views image $I_k$ and so the observations have density $f_k$ corresponding to $I_k$. The outcome of other actions for this hypothesis are explained similarly. An analogous description holds for outcomes of actions under $H_i$ when $i > W$.

The relative entropies for the various combinations of hypotheses pairs $(i \ne j)$ and actions are as follows. The expressions are self-explanatory.

\begin{align}
 \label{eqn:KL1}
 (i) \hspace{1cm}& \text{ $i \le W$, $j \le W$:}\\
 &D(q_{i}^{a} \Vert q_{j}^{a}) =
\begin{cases}
 D(f_{k} \Vert f_{l})& \text{$a=i$}\\
 D(f_{l} \Vert f_{k})& \text{$a=j$}\\
 0& \text{$a\ne i$, $a \ne j$}.
\end{cases}\\
\label{eqn:KL2}
 (ii) \hspace{1cm} & \text{ $i \le W$, $j=i+W$:}\\
&D(q_{i}^{a} \Vert q_{j}^{a}) =
\begin{cases}
 D(f_{k} \Vert f_{l})& \text{$a=i$}\\
 D(f_{l} \Vert f_{k})& \text{$a\ne i$}.
\end{cases}\\
\label{eqn:KL3}
(iii) \hspace{1cm} &  \text{ $i \le W$, $j > W$, $j \ne i+W$:}\\
&D(q_{i}^{a} \Vert q_{j}^{a}) =
\begin{cases}
 0& \text{$a=i$}\\
 0& \text{$a=j-W$}\\
 D(f_{l} \Vert f_{k})& \text{$a\ne i$, $a\ne j-W$}.
\end{cases}
\end{align}
$(iv)$  { For $i > W$, the expressions for $j > W$, $j = i -W$, or $j < W$ but $j \ne i-W$ are similar to $(i)$, $(ii)$, and $(iii)$ above, respectively, but with $f_{k}$ and $f_{l}$ interchanged.}

We now identify the structure of $\lambda_{i}$ and $D_{i}$ for $i = 1, 2, \ldots, 2W$.

\begin{proposition}
\label{prop:optimum Di for Case 2}
Let $W \ge 3$. Let $i \leq W$. For the setting of Case 2, the optimum $\lambda_{i}$ and $D_{i}$ of (\ref{eqn:optimal_lambda}) and (\ref{eqn:D_i}), respectively, are as follows.
If $D(f_{k}\Vert f_{l}) > D(f_{l}\Vert f_{k}) / (W-1)$ then
\begin{align}
 \nonumber \lambda_{i}(i)&=\frac{(W-3)D(f_{l}\Vert f_{k})}{(W-1) D(f_{k}\Vert f_{l})+(W-3) D(f_{l} \Vert f_{k})},\\
 \nonumber\lambda_{i}(j)&=\frac{D(f_{k}\Vert f_{l})}{(W-1) D(f_{k}\Vert f_{l})+(W-3) D(f_{l} \Vert f_{k})} \qquad \forall j \ne i,~ and ~\\
\label{eqn: optimum Di case 2} D_{i} &= \frac{(W-2)D(f_{k}\Vert f_{l})D(f_{l}\Vert f_{k})}{(W-1) D(f_{k}\Vert f_{l})+(W-3) D(f_{l} \Vert f_{k})}.
\end{align}
If $D(f_{k}\Vert f_{l}) \le D(f_{l}\Vert f_{k}) / (W-1)$ then
\begin{align*}
 \lambda_{i}(i)&=0,~ \lambda_{i}(j)=\frac{1}{(W-1)} ~ \forall j \ne i, ~ and ~ D_{i} = \frac{D(f_{l}\Vert f_{k})}{(W-1)}. %\label{eqn: optimum lambda 2}
\end{align*}
For $i > W$, $\lambda_{i}$ and  $D_{i}$ have the same structure as above, but with $f_{l}$ and $f_{k}$ interchanged.
\end{proposition}

For a proof, see Appendix \ref{proof:proof of optimum Di}.

\section{Proposal for a neuronal dissimilarity index}
\label{sec:ProposalForANeuronalMetric}

We now apply the results obtained in the previous section to the data from the experiments of Sripati and Olson \cite{ref:201001JNS_SriOls}. The visual search experiments of Sripati and Olson \cite{ref:201001JNS_SriOls} on human subjects correspond closely with Case 2 of the previous section. Similar to Case 2, the subjects in the experiments had no prior information on which of the two images $I_k$ and $I_l$ was the oddball image and which the distractor. But different from Case 2, the subjects in the experiments had to learn about the images $I_{k}$ and $I_{l}$ on-the-go, while in Case 2 we assume that the subject knows that the oddball and distractor images come from the set $\{ I_k, I_l\}$. A more accurate modeling that takes the learning aspect into account is work in progress. Here, we shall proceed with the Case 2 model.

Recall  that $T$ is the slot duration during which the subject focuses attention on a particular image.  First, we calculate the relative entropy $D(f_{k}\Vert f_{l})$ when $f_{k}$ and $f_{l}$ are densities of vector Poisson point processes of duration $T$ with rates $\mathbf{R}_{k} = (R_{k}(1), R_{k}(2), \ldots, R_{k}(d))$ and  $\mathbf{R}_{l} = (R_{l}(1), R_{l}(2), \ldots, R_{l}(d))$. Under the assumption that the neurons fire independently with the specified rates, the relative entropy decomposes into a sum:

\begin{align*}
 D &\left(\mu_{\mathbf{R}_{k},T}\Vert \mu_{\mathbf{R}_{l},T}\right)  = E_{\mu_{\mathbf{R}_{k},T}} \left[\log \frac{d\mu_{\mathbf{R}_{k},T}}{d\mu_{\mathbf{R}_{l},T}} \right]\\
& = \sum_{m=1}^{d} E_{\mu_{\mathbf{R}_{k}(m),T}} \left[\log \frac{d\mu_{\mathbf{R}_{k}(m),T}}{d\mu_{\mathbf{R}_{l}(m),T}} \right]\\
& = T\sum_{m=1}^{d}\left[R_{k}(m) \log \left( \frac{R_{k}(m)}{R_{l}(m)} \right) -R_{k}(m)+R_{l}(m)\right],
\end{align*}
where the term within square brackets in the last summation is the relative entropy of the Poisson point processes with rate $R_{k}(m)$ taken with respect to another such process with rate $R_{l}(m)$.

In Case 2, if the number of locations $W =6$, if $I_{k}$ is the oddball image, if $I_l$ is the distractor image, and if $D\left(\mu_{\mathbf{R}_{k},T}\Vert \mu_{\mathbf{R}_{l},T}\right) > D\left(\mu_{\mathbf{R}_{l},T}\Vert \mu_{\mathbf{R}_{k},T}\right) / (W-1)$ which is the case when $D\left(\mu_{\mathbf{R}_{l},T}\Vert \mu_{\mathbf{R}_{k},T}\right)$ is close to $D\left(\mu_{\mathbf{R}_{k},T}\Vert \mu_{\mathbf{R}_{l},T}\right)$, then from Proposition \ref{prop:optimum Di for Case 2} we have
\begin{align}
{D_{kl}} =   \frac{4D(\mu_{\mathbf{R}_{k},T} \Vert \mu_{\mathbf{R}_{l},T})D(\mu_{\mathbf{R}_{l},T} \Vert \mu_{\mathbf{R}_{k},T})}{5D(\mu_{\mathbf{R}_{k},T} \Vert \mu_{\mathbf{R}_{l},T})+3D(\mu_{\mathbf{R}_{l},T} \Vert \mu_{\mathbf{R}_{k},T})}.
\end{align}
Similarly, if $I_{l}$ is the oddball image, if $I_k$ is the distractor image, and if $D\left(\mu_{\mathbf{R}_{l},T}\Vert \mu_{\mathbf{R}_{k},T}\right) > D\left(\mu_{\mathbf{R}_{k},T}\Vert \mu_{\mathbf{R}_{l},T}\right) / (W-1)$, we have
\begin{align}
{D_{lk}} =   \frac{4D(\mu_{\mathbf{R}_{l},T} \Vert \mu_{\mathbf{R}_{k},T})D(\mu_{\mathbf{R}_{k},T} \Vert \mu_{\mathbf{R}_{l},T})}{5D(\mu_{\mathbf{R}_{l},T} \Vert \mu_{\mathbf{R}_{k},T})+3D(\mu_{\mathbf{R}_{k},T} \Vert \mu_{\mathbf{R}_{l},T})}.
\end{align}
Let us normalize $D_{kl}$ per unit time and per neuron and denote it $\tilde{D}_{kl}$:
\begin{align}
\label{eqn:neuronal metric Di tilde}
 \tilde{D}_{kl} = \frac{1}{dT}D_{kl}.
\end{align}

The subset of experimental data gathered by Sripati and Olson that we use in our analysis consisted of the following.

1) Neuronal firing rate vectors were obtained from the IT cortex of rhesus macaque monkeys for twenty four images. The number of neurons ranged from 78 to 174, the variation was due to experimental constraints. But the sets of neurons tapped were identical for images that were to be paired in the decision time experiments on human subjects, which we describe next.

2) Decision time statistics for detection of the oddball image were obtained from experiments on human subjects. For oddball image $I_k$ and distractors $I_l$, data was collected as follows. Six subjects participated and each was shown twelve stimuli. In each stimulus, the oddball location was picked uniformly at random from the $W=6$ locations. The decision time were averaged across various subjects and across stimuli instances to get $s(k,l)$. The first argument $k$ stands for the oddball image $I_k$.

Recall from Case 2 that $H_i$, when $i \leq W = 6$, is the hypothesis that the oddball image is $I_k$ and the distractor images are $I_l$. Taking a cue from Theorem \ref{thm:asymptotic optimality of sluggish procedure A}, assuming a sufficiently stringent error tolerance vector of $(\alpha, \alpha, \ldots, \alpha)$ for $\alpha$ sufficiently small, and assuming nearly optimal decision making, we predict that
\[
  E_i [C(\pi)] \approx \frac{\log (1/\alpha)}{D_{kl}},
\]
{where $C(\pi)$ models the total decision time, the sum of sensing delay and switching delays.}
Averaging across $i = 1, \ldots, W$, i.e., averaging across all those stimuli where $I_k$ is the oddball image and $I_l$ is the distractor, we get
\begin{eqnarray*}
  E[C(\pi) ~|~ I_k \text{ is the oddball and } I_l \text{ is the distractor} ]
  & \approx & \frac{\log (1/\alpha)}{D_{kl}} \\
  & = & \frac{(1/dT) \log (1/\alpha)}{\tilde{D}_{kl}},
\end{eqnarray*}
or in other words
\[
  s(k,l) \cdot \tilde{D}_{kl} \approx constant.
\]
For $i > W=6$, one similarly has
\[
  s(l,k) \cdot \tilde{D}_{lk} \approx constant.
\]
This naturally leads to the proposal
\begin{equation}
  \label{eqn:diff-proposal}
  \textsf{diff}(\textbf{R}_k, \textbf{R}_l) = \tilde{D}_{kl}.
\end{equation}

\subsection{Correlation study}
\label{subsec:correlationstudy}

\begin{figure}[ht]
  \centering
  \includegraphics[scale=0.55]{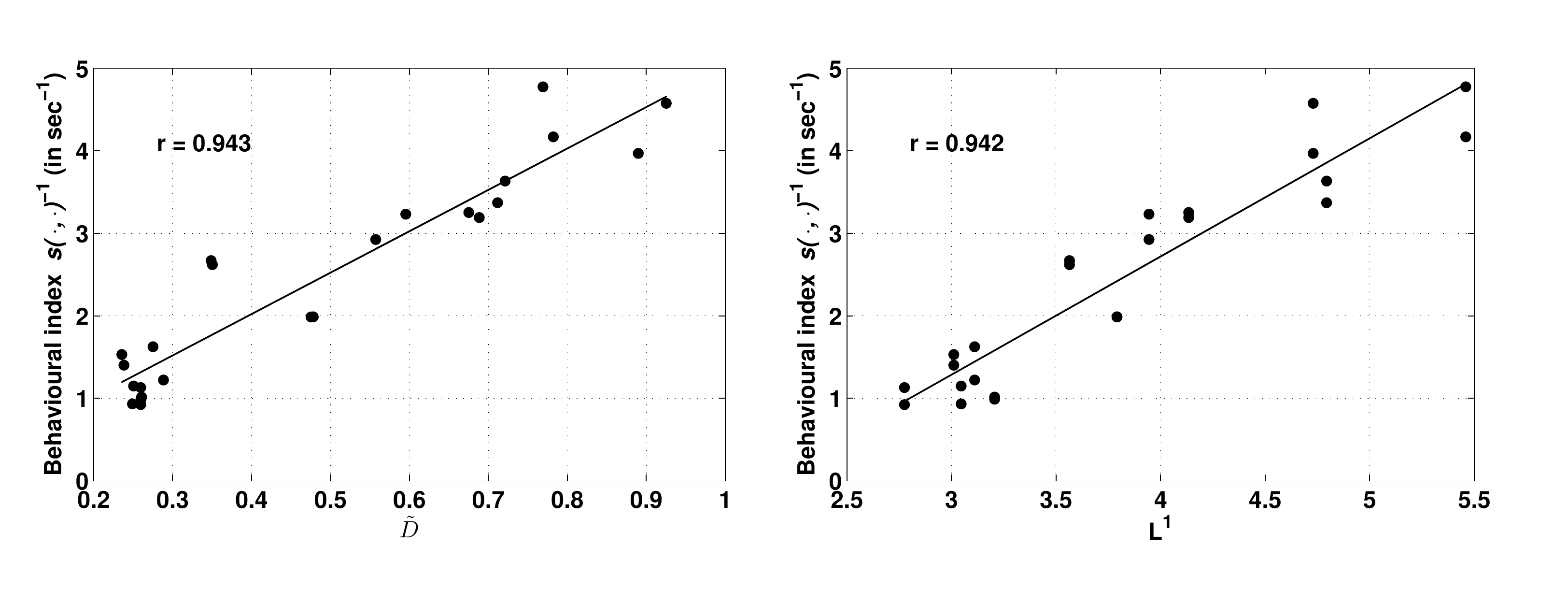}
  \caption{The observed behavioural dissimilarity index versus the proposed neuronal dissimilarity index $(\tilde{D})$ and the $L^{1}$-neuronal index.}
  \label{fig:Corr1}
\end{figure}

The {\em behavioural dissimilarity index} for an ordered pair of images $(k,l)$ is the inverse $s(k,l)^{-1}$ of the average decision time $s(k,l)$, and gives an indication of the speed of discrimination. In Figure \ref{fig:Corr1}, we plot the behavioural dissimilarity index $s(k,l)^{-1}$ against the proposed neuronal dissimilarity index $\tilde{D}_{kl}$ and against the $L^{1}$ dissimilarity index for various ordered pairs $(k,l)$. We observe a strong correlation of $0.94$ for $\tilde{D}_{kl}$ which is the same as the correlation between the behavioural dissimilarity index and the $L^{1}$ distance $|| \mathbf{R}_k - \mathbf{R}_l||_{1}$.

\begin{figure}[ht]
 \centering
 \includegraphics[scale=0.55]{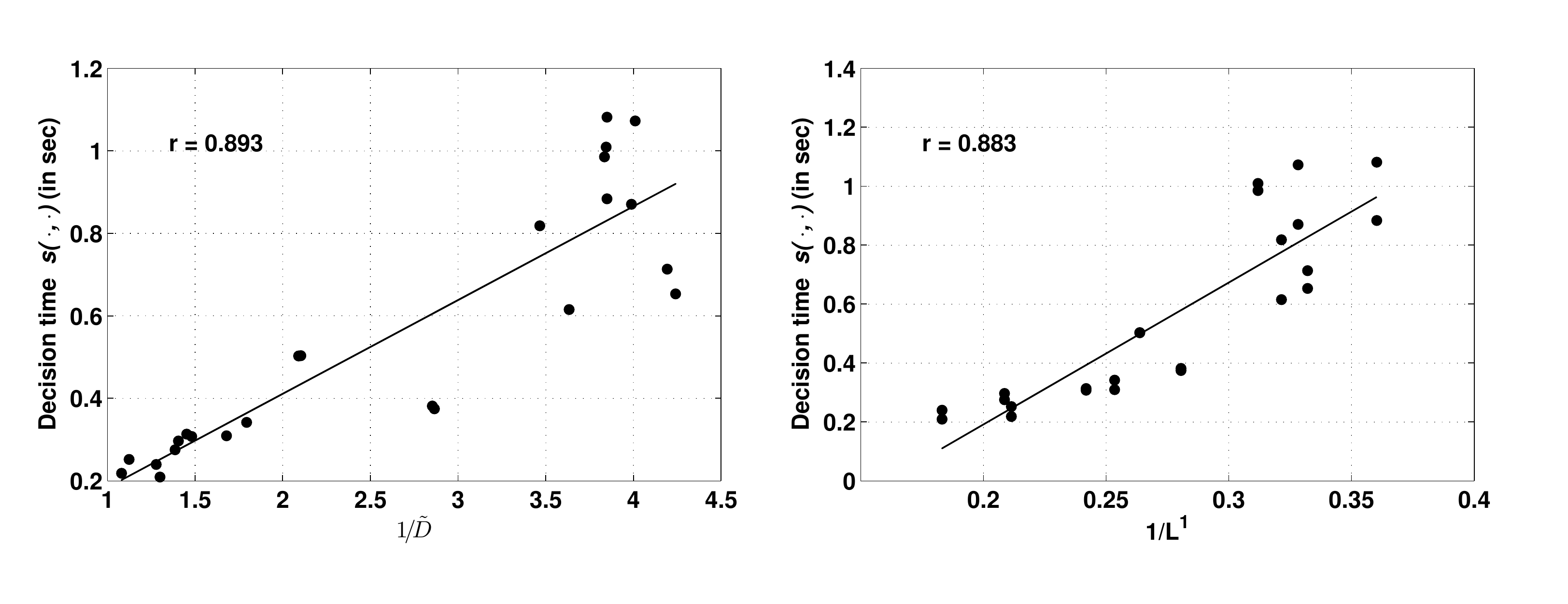}
\caption{The observed decision time versus the inverse of the proposed neuronal index $(1 / \tilde{D})$ and the inverse of the $L^{1}$-neuronal index.}
\label{fig:detection_delay_vs_inverse_neu_metric}
\end{figure}

Now that we have discovered that our proposed neuronal dissimilarity index and the $L^1$ index are both equally well-correlated with the behavioural dissimilarity index, it is natural to ask if there is some basis to choose one over the other. The point that our proposed dissimilarity index has a ``microscopic basis'' (grounded in decision theory and based on the ASHT framework) that explains the ``macroscopic observations'' (speed of discrimination) is certainly in our favour. But there are other related dissimilarity indices such as relative entropy (KL) and Chernoff entropy that have similar correlation with the behavioural dissimilarity index. Table \ref{table:correlation of different neuronal metrics 2} summarises the correlations (second column) along with their $p$-values (third column). It is therefore natural to ask if a finer examination of the experimental data can help us identify the ``best'' among these neuronal indices. We shall pursue this in the next subsection and shall propose a method to rank order the indices in terms of their ability to explain the experimental data.

\begin{table}[t]
\caption{Correlation with Different Neuronal Dissimilarity Indices} % title of Table
\centering % used for centering table
\begin{tabular}{|c|c|c||c|c|c|} % centered columns (4 columns)
\hline\hline %inserts double horizontal lines
Information Measure & Correlation ($1/s$ vs. Neuronal index) & $p$-value & Correlation ($s$ vs $(\text{Neuronal index})^{-1}$)& $p$-value\\  % inserts table
%heading
\hline % inserts single horizontal line
&&&&\\
Proposed $\tilde{D}$ & 0.94 & $5.2 \times 10^{-12}$ & 0.89 & $4.3 \times 10^{-09}$\\
KL & 0.93 & $8.5 \times 10^{-11}$ & 0.90 & $3.1 \times 10^{-09}$\\
Chernoff & 0.94 & $7.8 \times 10^{-12}$ & 0.88 & $2.1 \times 10^{-08}$\\
$L^1$ & 0.94 & $6.1 \times 10^{-12}$ & 0.88 & $1.1 \times 10^{-08}$\\ % inserting body of the table
[1ex] % [1ex] adds vertical space
\hline %inserts single line
\end{tabular}
\label{table:correlation of different neuronal metrics 2} % is used to refer this table in the text
\end{table}

\vspace*{.1in}

A more basic question, and one that is motivated by our expectation that $s(k,l) \cdot \textsf{diff}(\mathbf{R}_k, \mathbf{R}_l) = constant$, is whether it is more appropriate to correlate $s(k,l)$ versus $\textsf{diff}(\mathbf{R}_k, \mathbf{R}_l)^{-1}$ as opposed to what is done in Figure \ref{fig:Corr1} which correlates $s(k,l)^{-1}$ versus $\textsf{diff}(\mathbf{R}_k, \mathbf{R}_l)$. Table \ref{table:correlation of different neuronal metrics 2} reports these correlations (fourth column) along with the corresponding $p$-values (fifth column), and Figure \ref{fig:detection_delay_vs_inverse_neu_metric} provides the correlation plot. The new correlations, though high, are lower than those reported in the second column.

We do not have a clear-cut answer on which of the two scatter plots --
\begin{equation}
  \label{eqn:which-correlation}
  (s(k,l), \textsf{diff}(\mathbf{R}_k, \mathbf{R}_l)^{-1}) \text{ or } (s(k,l)^{-1}, \textsf{diff}(\mathbf{R}_k, \mathbf{R}_l))
\end{equation}
-- and the corresponding correlations is more appropriate. However, recall that Pearson's test for rejecting the null hypothesis that a bivariate normal has independent components is that the correlation statistic $r$ arising from independent and identically distributed samplings of the bivariate normal has $|r|$ exceeding a threshold. Given that $s(k,l)$ is the arithmetic mean of $n=72$ experimentally measured decision time, when centred and scaled, $s(k,l)$ is likely to be closer to normal than its inverse. We therefore believe the correlation of $(s(k,l), \textsf{diff}(\mathbf{R}_k, \mathbf{R}_l)^{-1})$, the one that leads to lower correlations, is more appropriate. The indicated $p$-values, shown in Figures \ref{fig:Corr1} and \ref{fig:detection_delay_vs_inverse_neu_metric} and in Tables \ref{table:correlation of different neuronal metrics} and \ref{table:correlation of different neuronal metrics 2}, are the probabilities that the correlation statistic equals or exceeds the indicated observed levels when the null hypothesis is true (independent components).

%%%%%%%%%%%%%%%%%%%%%%%%%%%%%%%%%%%%%%%%%%%%%%%%%%%%%%%%%%%%%%%%%%%%%%%%%%%%%%%%%%%%%%%%%%%%%%%%%%%%%%%%%%%%
\subsection{Model testing via three ``equality of means'' tests}
\label{subsec:equalityOfMeans}

In Section \ref{Introduction}, we posed the question of identifying a suitable \textsf{diff} function that satisfies
\begin{align}
 \label{eqn:suitable diff function}
 s(k,l) \cdot \textsf{diff}(\mathbf{R}_{k},\mathbf{R}_{l}) = constant.
\end{align}
In the previous section, we modeled visual search as an ASHT problem and proposed the $\textsf{diff}$ given in (\ref{eqn:diff-proposal}), denoted $\tilde{D}$. However, we also saw that the candidates $L^1$,  Chernoff entropy, relative entropy, and $\tilde{D}$, all yielded high correlation with the behavioural dissimilarity index. We now address the question of which of these dissimilarity indices best explain the data.

%We now attempt a finer analysis to see if the experimental data can help us rank these various $\textsf{diff}$ proposals.

Our methodology is as follows. Consider a fixed $\textsf{diff}(\mathbf{R}_k, \mathbf{R}_l)$ function. Let us test the new null hypothesis:
\[
(H_{0}) \quad : \quad E[C(\pi) ~|~ I_k \text{ is the oddball and } I_l \text{ is the distractor} ] \cdot \textsf{diff}(\mathbf{R}_k, \mathbf{R}_l) = constant,
\]
where $C(\pi)$ is the decision time for a fixed error tolerance on the ordered image pair $(I_k,I_l)$.
The decision time data across subjects and across multiple stimuli that have $I_k$ as the oddball and $I_l$ as the distractor images constitute one group associated with the ordered pair $(k,l)$. $H_0$ hypothesises that the $\textsf{diff}$-scaled means is constant across groups. Let us identify the $\textsf{diff}$ indices for which the corresponding null hypothesis is accepted for a desired significance level. If the test passes for $\textsf{diff}(\mathbf{R}_k, \mathbf{R}_l) = \tilde{D}_{kl}$, then there is significant evidence that the data is well-explained by our theory.

To perform this test, we must do the following for each $\textsf{diff}$ candidate.
\begin{itemize}
  \item Identify a test statistic $\mathcal{T}(\textsf{diff})$ for testing equality of means of the $\textsf{diff}$-scaled decision times. Note each $\textsf{diff}$ leads to a separate hypothesis test.
  \item Accept or reject the corresponding null hypothesis for a desired level of significance.
\end{itemize}

Let $\tau_{k,l}(j)$ be the $j$th sample in the group indexed by $(k,l)$. Let $n$ denote the common number of samples in each group, and let $g$ be the number of groups. The experimental data of Sripati and Olson had 24 groups and 72 samples per group; $n=72$ and $g=24$. The number of samples in each group was identical.

\subsubsection{Test 1 - Oneway ANOVA}: \label{subsubsec:onewayAnova}
The one-way analysis of variance (ANOVA) statistic \cite{ref:CasellaBerger_StatisticalInference} is often used to test equality of means across groups when the samples are Gaussian and when the variances across groups are the same. This test is known to be robust to the Gaussian assumption. It is also known to be robust to the equality of variances assumption so long as the number of samples is the same across groups \cite[p.243]{glass1972consequences}. As we will soon see, we neither have Gaussianity nor equality of variances across groups. But since the number of samples is the same across the groups, we may still use the oneway ANOVA test.

Let $T_{k,l}(j) := \tau_{k,l}(j) \cdot \textsf{diff}(\mathbf{R}_k, \mathbf{R}_l)$. Write the sample means, the mean across groups, and the pooled variance as follows.
\begin{eqnarray*}
  \bar{T}_{k,l} & = & \frac{1}{n} \sum_{j=1}^{n} T_{k,l}(j), \\
  \bar{\bar{T}} & = & \frac{1}{g} \sum_{(k,l)} \bar{T}_{k,l}, \\
  S_p^2 &=& \frac{1}{g(n-1)} \sum_{(k,l)} \sum_{j=1}^n (T_{k,l}(j) - \bar{T}_{k,l})^2.
\end{eqnarray*}
Note that these depend on the $\textsf{diff}$ index under consideration. The oneway ANOVA test \cite[p.533]{ref:CasellaBerger_StatisticalInference} is as follows: Reject $H_0$ (associated with the $\textsf{diff}$ under consideration) if
\[
  \mathcal{T}(\textsf{diff}) := \frac{\displaystyle  \sum_{(k,l)} n \left( \bar{T}_{k,l} - \bar{\bar{T}} \right)^2}{S_p^2} > (g-1) F_{g-1, g(n-1), \alpha},
\]
where $\alpha$ is the desired significance level and $F_{g-1, g(n-1), \alpha}$ is the corresponding threshold\footnote{The threshold at which the cdf of the $F$-distribution with $(g-1, g(n-1))$ degrees of freedom equals $1-\alpha$.}.

Look at the second and third columns of Table \ref{table:GLRT_values}. The first row contains the value of the ANOVA statistic with $\textsf{diff}(\mathbf{R}_k, \mathbf{R}_l) = \tilde{D}_{kl}$ and the corresponding $p$-value. The $p$-value is so small that we must summarily reject the null hypothesis $H_0$ associated with $\textsf{diff} = \tilde{D}$ (for, say, a typical significance level of 5\%). The situation is the same for the other dissimilarity indices, as can be seen from the remaining rows of Table \ref{table:GLRT_values}. In each test, the null hypothesis is rejected for, say, the typical 5\% significance level.

Observe that the values of $\mathcal{T}(\tilde{D})$, $\mathcal{T}(\text{KL})$, and $\mathcal{T}(\text{Chernoff})$ are close to each other while $\mathcal{T}(L^1)$ is significantly larger. This suggests that one could use $\mathcal{T}(\cdot)$ to rank the different dissimilarity measures in their ability to explain the observed data. The oneway ANOVA statistic suggests the ranking
\begin{equation}
  \label{eqn:ordering}
  \tilde{D} > \text{KL} > \text{Chernoff} > L^{1}.
\end{equation}
We shall return to this observation after trying out two other refinements of the equality of means test.

\subsubsection{Equality of Means for Gamma Distributions}
We began with the oneway ANOVA statistic because it is known to be robust to the Gaussian assumption. We checked for Gaussianity anyway. Lilliefor's test for Gaussianity is a variation on the Kolmogorov-Smirnov test when the null hypothesis does not specify the parameters of the Gaussian distribution. None of the 24 groups of data passed the test of Gaussianity at the 5\% significance level.

\begin{figure}[t]
\centering
\includegraphics[scale=0.55]{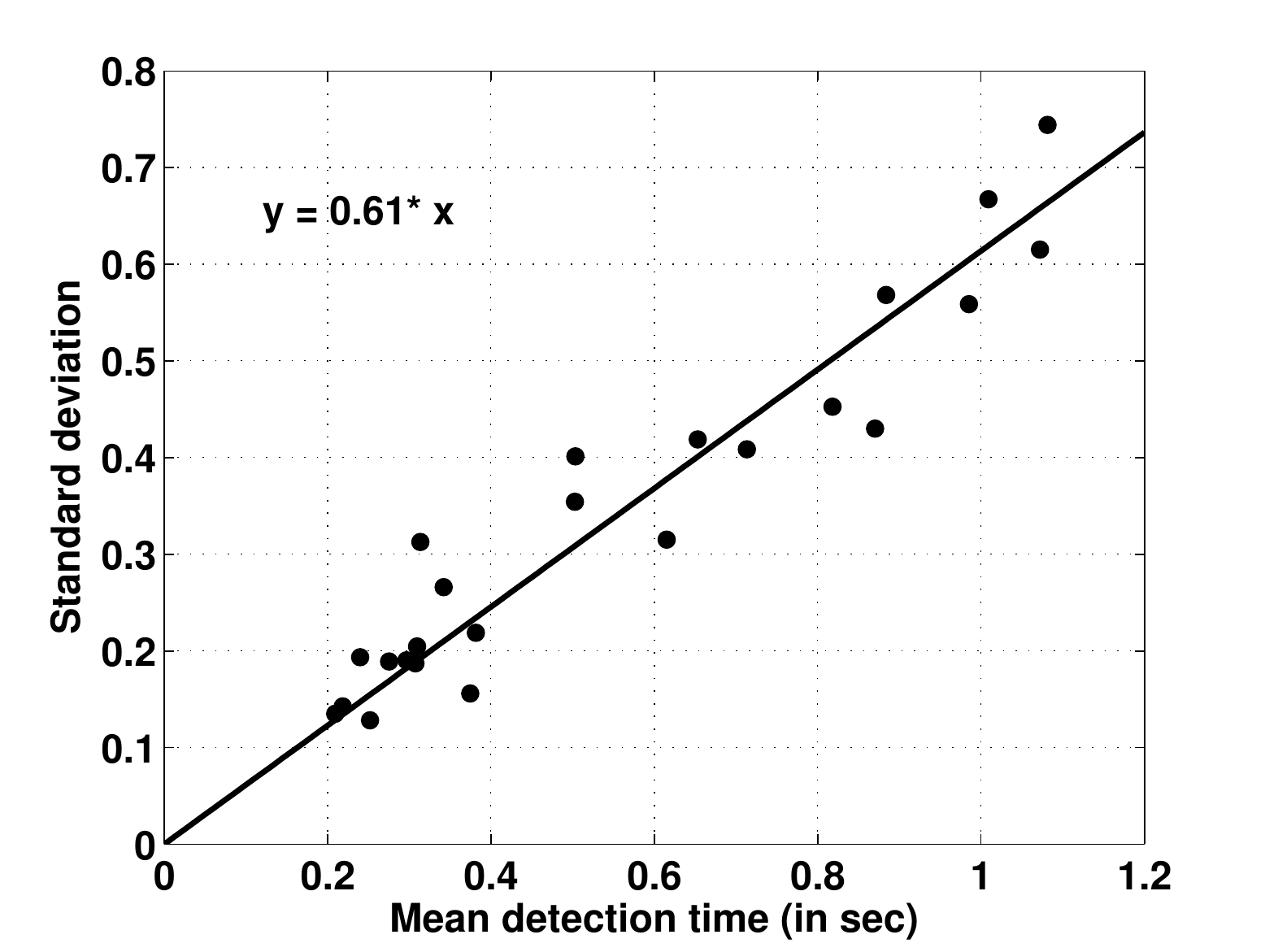}
\caption{Standard deviation of the decision times versus mean decision times, across image pairs.}
\label{fig:mean_vs_std_search_time}
\end{figure}

We next looked for features in the data that may suggest other distributions. First, the decision times are positive random variables. Next, Figure \ref{fig:mean_vs_std_search_time} shows the standard deviation versus the mean decision time for the 24 image pairs. Observe the linear relation, with $y=0.61x$ being the best linear fit. A class of distributions on $\mathbb{R}_{+}$ whose standard deviation is a linear function of its mean is the family of Gamma distributions with a fixed shape parameter. The Gamma density with shape parameter $s$ and scale $m$ is
\[
  (m \Gamma(s))^{-1}(x/m)^{s-1} e^{-x/m}.
\]
{The mean is $ms$ and the standard deviation is $m\sqrt{s}$ so that standard deviation to mean ratio is $1/\sqrt{s}$. The slope of $0.61$ in Figure \ref{fig:mean_vs_std_search_time} suggests a shape parameter of $1/(0.61)^2 = 2.7$. The Kolmogorov-Smirnov test on the data against Gamma distributions with shape parameter $2.7$ and mean set to the sample mean accepts 18 of the 24 image pairs and rejects 6 out of 24 image pairs at 5\% significance level. This suggests that Gamma is a better fit to the data than Gaussian.}

\begin{figure}[t]
\centering
\includegraphics[scale=0.55]{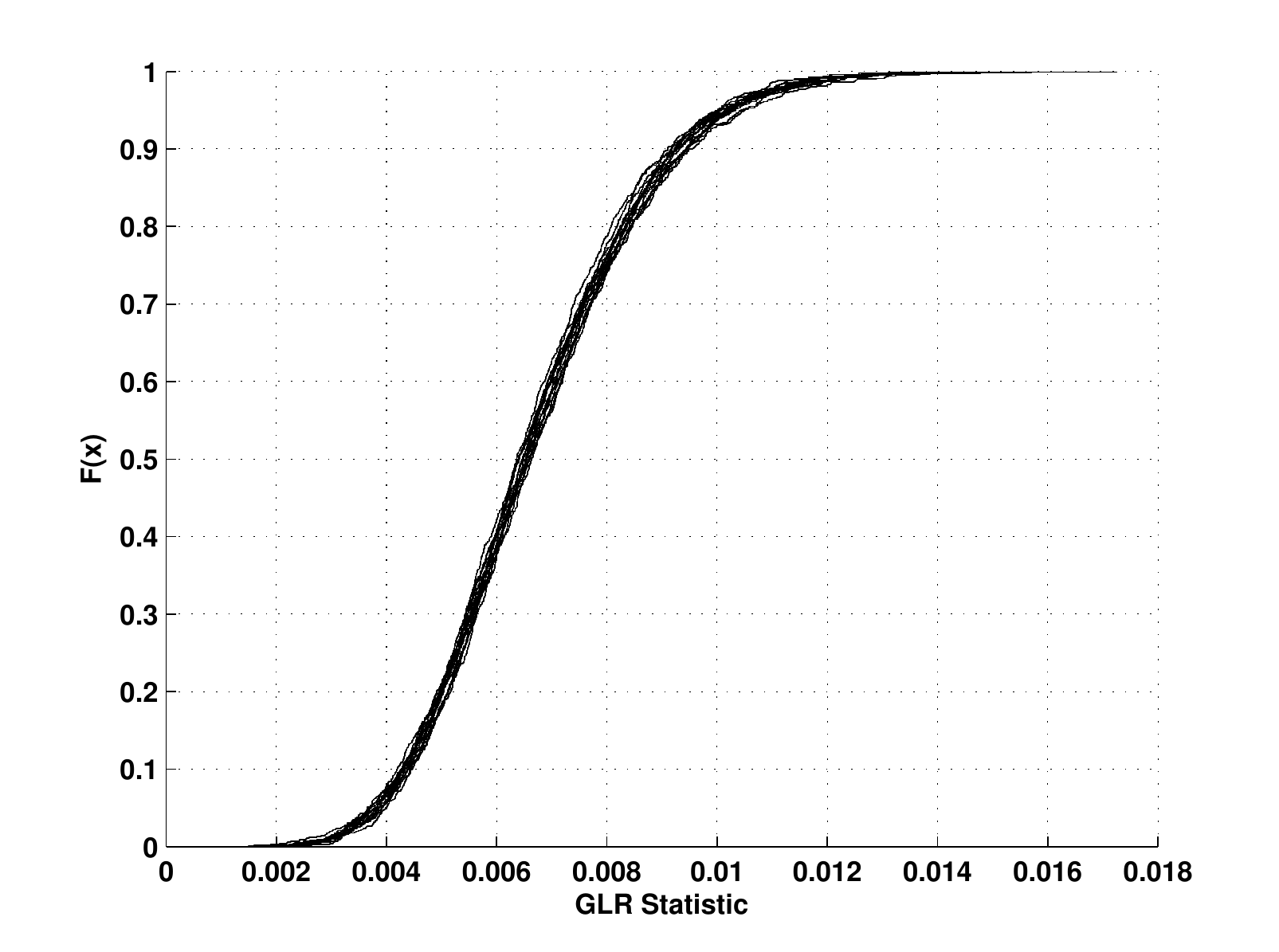}
\caption{Superposition of CDF of GLRT statistic under the Gamma assumption for the equality of means test. Each CDF consists of 1000 sample points. Each CDF corresponds to a random instance of mean and shape parameter. Mean was uniformly sampled from [0.2,1.2]. Shape parameter was uniformly sampled from [2,5]. Each sample point consisted of 24 groups, and 72 samples per group, same as that for the experimental data for decision times. The indicated intervals for mean and shape were based on the experimental data.}
\label{fig:gamma_GLR_CDF}
\end{figure}

We conducted a generalised likelihood ratio test (GLRT) for equality of means under the Gamma assumption, and under a constant shape parameter assumption. This corresponds to an ``equality of scales'' test; Shiue et al. \cite{ref:ShiueBainEngel_Equality_Gamma_1988} suggest a statistic analogous to the oneway ANOVA but for Gamma distributions. In Figure \ref{fig:gamma_GLR_CDF} we plot the CDF of the GLRT statistic under the null hypothesis and under equal mean and equal shape parameter assumptions. Note that the CDF of this statistic is robust to the shape parameter.

In column 4 of Table \ref{table:GLRT_values} we provide the GLRT statistics for the decision time data. If we compare the GLRT statistics from Table \ref{table:GLRT_values} against GLRT CDF in Figure \ref{fig:gamma_GLR_CDF}, we observe that the statistics is well beyond the $5\%$ significance point. Again, we must summarily reject each of the equality of means hypotheses. Indeed, each GLRT statistic is off the chart in Figure \ref{fig:gamma_GLR_CDF}. However, direct ordering of the statistics suggests the ranking (\ref{eqn:ordering}), the same as that obtained with ANOVA.

\subsubsection{Equality of Means for Gamma Distributions with a fixed and known shape parameter}

Consider the setting where the shape parameter is known to be $s$ across the groups. The GLRT for equality of means under this setting can be straightforwardly shown to be $s \log (\text{AM} / \text{GM})$ where AM and GM are the arithmetic and geometric means across groups defined as follows:
\[
  \text{AM} = \bar{\bar T} = \frac{1}{g} \sum_{(k,l)} \bar{T}_{k,l} \mbox{ and } \text{GM} = \left( \prod_{(k,l)} \bar{T}_{k,l} \right)^{1/g}.
\]
The last column of Table \ref{table:GLRT_values} shows this statistic for the various dissimilarity measures. Yet again, the statistics are off the chart (CDF not plotted) with so small $p$-values that the null hypothesis of equality of means must be rejected. Once again, direct ordering of statistics suggests the ranking (\ref{eqn:ordering}).

\subsubsection{A lesser goal - Ranking}

We saw that all three equality of means tests reject all four dissimilarity measures. In retrospect, this might have been anticipated. If the test associated with $\tilde{D}$ had passed, that would have been a spectacular confirmation of our theory, which we really did not expect due to the crudeness of our modeling. Nevertheless, the equality of means test statistic provides a means to check which of the four dissimilarity measures best explains the data.

A little thought will inform us that all three equality of means tests check how clustered the sample means are, across various groups. This is clearest in the third equality of means test (for Gamma distributions with a known and common shape parameter across groups) where the statistic is a monotone function of the ratio AM/GM. The test passes if AM/GM is small, that is if the group means are close to each other, and fails if it is large.

The columns of Table \ref{table:GLRT_values} corresponding to each test statistic suggest that the data points are most clustered under $\tilde{D}$ scaling and least clustered with $L^1$ scaling. The ranking is as in (\ref{eqn:ordering}). 

Let us note $\tilde{D}$, KL, and Chernoff are close to each other, and $L^1$ a distant fourth. Indeed, the vector $(\bar{T}_{(k,l)}/\bar{\bar{T}})_{(k,l)}$ associated with the $L^1$ dissimilarity measure {\em majorised} (\cite[Defn.A.1]{marshall2010inequalities}) the other three. There was no such ordering among $\tilde{D}$, KL, and Chernoff.

\begin{table}[ht]
\caption{Equality of means test. Various Statistics} % title of Table
\centering % used for centering table
\begin{tabular}{|l|c|c|c|c|} % centered columns (4 columns)
\hline\hline %inserts double horizontal lines
 $\textsf{diff}$ & ANOVA statistic & ANOVA $p$-values & Gamma GLR &  $s$log(AM/GM)\\  % inserts table
%heading
\hline % inserts single horizontal line
& & & &\\
 Proposed $\tilde{D}$ & $06.30$ & $9.35 \times 10^{-19}$ & $0.0533$ & $0.0600$ \\
       KL & $06.68$ & $2.88 \times 10^{-20}$ & $0.0561$ & $0.0633$ \\
 Chernoff & $06.74$ & $1.61 \times 10^{-20}$ & $0.0663$ & $0.0756$ \\
  $L^{1}$ & $24.00$ & $3.42 \times 10^{-87}$ & $0.1652$ & $0.2061$ \\
\hline %inserts single line
\end{tabular}
\label{table:GLRT_values} % is used to refer this table in the text
\end{table}

\section{Discussion}
\label{sec:Conclusion}

We modelled the visual search task of Sripati and Olson \cite{ref:201001JNS_SriOls} as an active sequential hypothesis testing problem (ASHT). We extended the ASHT results of Chernoff \cite{ref:195909AMS_Che} to the case with switching costs. We showed that adding switching costs does not affect the asymptotic growth rate of the total cost.

The ASHT model suggests a dissimilarity index between pairs of images. The inverse of the asymptotic growth rate of the total cost in the ASHT model is proposed as a dissimiliarity index between pairs of images. We derived expressions for computing the proposed dissimilarity index for the specific search task considered by Sripati and Olson \cite{ref:201001JNS_SriOls}. The proposed dissimilarity index is a function of the neuronal firing rates elicited by the images in the infero temporal cortex of macaque monkeys.

Correlation study indicated that the proposed index is as good as $L^1$ and other dissimilarity measures such as the Chernoff entropy and the relative entropy (KL). Equality of means testing indicated that the equality of means hypothesis should be rejected, and this can be done with overwhelming confidence. 

Equality of means testing procedure is perhaps a rather stringent test. What would be an appropriate test if, say, we can leave one group out? Does our proposed neuronal dissimilarity index pass such a less stringent test? Can we leave two groups out? Which two? We do not yet have a principled way to address these questions and instead decided to stick to the strictest test.

The statistics associated with the equality of means testing, however, suggested a ranking of the dissimilarity measures. We proposed three different statistics. Each measures the spread across groups of the group sample means. One of them is the familiar AM/GM ratio. The ranking was consistent across the three different statistics. Our proposed index was ranked first, relative entropy (KL divergence) and Chernoff entropy were a close second and third, and $L^1$ was a somewhat distant fourth.

The decision times were tested for the Gamma distribution and the test passed for two-third of the groups. The shape parameter for the distributions of delay, estimated via the method of moments, was close to 3. 

{In our work, we took only valid trials, i.e., those where the decisions made by the subjects were correct. We also assumed that the error probability tolerance were the same across subjects. It would be interesting to model speed-accuracy tradeoffs and see how they vary across individuals. It would also be interesting to explore how they vary for a single subject under different incentive settings.}

Extension of ASHT to the case when no prior information is available about the images, where the subject has to actively learn $\mathbf{R}_k$ and $\mathbf{R}_l$ on-the-fly, is an interesting learning problem that is currently under study.

%%%%%%%%%%%%%%%%%%%%%%%%%%%%%%%%%%%%%%%%%%%%%%%%%%%%%%%%%%%%%%%%%%%%%%%%%%%%%%%%%555
\appendices
%%%%%%%%%%%%%%%%%%%%%%%%%%%%%%%%%%%%%%%%%%%%%%%%%%%%%%%%%%%%%%%%%%%%%%%%%%%%%%%%%%%%

%%%%%%%%%%%%%%%%%%%%%%%%%%%%%%%%%%%%%%%%%%%%%%%%%%%%%%%%%%%%%%%%%%%%%%%%%%%%%%%%%%%%
\section{Properties of log-likelihood ratio processes under $\pi_{SA}(L,\eta)$}
\label{sec:app:PropertiesOfLLR}

We will now show some desirable properties of the log-likelihood ratio processes under the policy $\pi_{SA}(L, \eta)$. These properties are analogous to those of classical sequential hypothesis testing, but their analyses are more involved because actions introduce 1) dependency in the log-likelihood ratio increments, and 2) the  increments are no longer identically distributed. The properties we will establish will be useful in forthcoming proofs.

Define $\Delta Z_{ji}(n) = Z_{ji}(n)-Z_{ji}(n-1)$. We then have $ \Delta Z_{ji}(n) = -\Delta Z_{ij}(n)$. Here, $\Delta Z_{ji}(n)$ is the increment in the process associated with the log-likelihood ratio of $H_{j}$ with respect to $H_{i}$ at time $n$. We now show that under Assumptions (I) and (IIb), and under policy $\pi_{SA}(L, \eta)$, the log-likelihood ratio processes are well behaved in the following sense: the log-likelihood ratio of the true hypothesis $H_{i}$ with respect to any other hypothesis $H_{j}$ has a positive drift. This will be made precise in Proposition \ref{prop:exponential decay LLR process}. Towards that, we first establish the following lemmas.

\begin{lemma}
 \label{lemma:likelihood ratio property conditioned on actions}
Assume (I) and (IIb).  Fix $i$, $j$ such that $j \ne i$. Let $a \in \mathcal{A}_{ij}$. We then have, for all $0 <s <1$,
\begin{align}
 \rho_{ij}^{a}(s) := E_{i}\left[e^{s\Delta Z_{ji}(n)} \vert A_{n} = a\right] < 1 \;\;\; \hfill \forall n.
\end{align}
\end{lemma}

\begin{IEEEproof}
The following sequence of inequalities hold:
\begin{align}
 \nonumber E_{i} &\left[e^{s\Delta Z_{ji}(n)} \vert A_{n} = a\right]\\
 \nonumber &= \int_{x \in \mathcal{X}} \left(\frac{q_{j}^{a}(x)}{q_{i}^{a}(x)}\right)^{s} q_{i}^{a}(x) dx\\
 \nonumber &=  \int_{x \in \mathcal{X}} \left(q_{j}^{a}(x)\right)^{s} \left(q_{i}^{a}(x)\right)^{1-s} dx \\
& < \left( \int_{x \in \mathcal{X}} q_{j}^{a}(x) dx \right)^{s} \left( \int_{x \in \mathcal{X}} q_{j}^{a}(x) dx \right)^{1-s} \label{eqn:Holders ineq}\\
 \nonumber& = 1.
\end{align}
The strict inequality in (\ref{eqn:Holders ineq}) follows from H\"{o}lder's inequality and the fact that $a \in \mathcal{A}_{ij}$ implies $q_{i}^{a}$ and $q_{j}^{a}$ are not linearly related.
\end{IEEEproof}

The above result was obtained by conditioning on the action $A_{n}$ to lie in the desirable set $\mathcal{A}_{ij}$. The result is independent of the underlying policy, because when conditioned on the current action $A_{n}$, the observation is independent of the policy.

Recall that  $\tilde{\pi}_{SA}(\eta)$ is the non-stopping variant of ${\pi}_{SA}(L, \eta)$. Further, recall from Assumption (IIb) that we have $\beta =  \min \left\{ \sum_{a \in \mathcal{A}_{ij}} \lambda_{k}(a) ~|~ 1 \leq i, j, k \leq M, ~ i \ne j \right\} > 0$. Now we show that, under Assumption (IIb) and policy $\tilde{\pi}_{SA}(\eta)$, a similar result holds, but without conditioning on the action $A_{n}$.
First, let us define
\begin{align}
 \rho_{ij}(s) :=  \eta \beta \left(\max_{a \in \mathcal{A}_{ij}}\rho_{ij}^{a}(s)\right)+ (1-\eta \beta).
\end{align}
The fact that $\rho_{ij}(s) < 1$ is evident from Lemma \ref{lemma:likelihood ratio property conditioned on actions}.

\begin{lemma}
 \label{lemma:likelihood ratio property under policy}
 Assume (I) and (IIb). Consider the  policy $\tilde{\pi}_{SA}(\eta)$. Fix $i$. We then have, for all $0<s<1$,
\begin{align*}
 E_{i} &\left[e^{s\Delta Z_{ji}(n)} \vert X^{n-1}, A^{n-1}\right] \le \rho_{ij}(s) < 1 \; \; \; \hfill \forall n, \forall j \ne i.
\end{align*}
\end{lemma}

\begin{IEEEproof}
The following sequence of inequalities hold as described after the last inequality.
\begin{align}
\nonumber E_{i} &\left[e^{s\Delta Z_{ji}(n)} \vert X^{n-1},A^{n-1}\right]\\
\nonumber &= E_{i} \left[ E_{i} \left[ e^{s\Delta Z_{ji}(n)} \vert X^{n-1},A^{n-1}, A_{n} \right] \vert X^{n-1},A^{n-1} \right]\\
\label{eqn:lemma rho_ij(s) equation 3} &=  \sum_{a \in \mathcal{A}} P_{i}(A_{n} =a\vert X^{n-1} A^{n-1}) E_{i} \left[e^{s\Delta Z_{ji}(n)} \vert A_{n} =a \right]\\
\nonumber &\le P_{i}(A_{n} \in \mathcal{A}_{ij}\vert X^{n-1} A^{n-1}) \max_{a \in \mathcal{A}_{ij}} E_{i} \left[e^{s\Delta Z_{ji}(n)} \vert A_{n} =a\right]\\
\label{eqn:lemma rho_ij(s) equation 4}& \hspace{0.5cm}+ (1-P_{i}(A_{n} \in \mathcal{A}_{ij}\vert X^{n-1} A^{n-1}))\\
\nonumber & \le \eta \beta \left(\max_{a \in \mathcal{A}_{ij}}\rho_{ij}^{a}(s)\right)+ (1-\eta \beta)\\
& < 1.
\end{align}
Equality (\ref{eqn:lemma rho_ij(s) equation 3}) holds because conditioned on $A_{n} = a$, $\Delta Z_{ij}(n)$ is independent of the remaining history. Inequality (\ref{eqn:lemma rho_ij(s) equation 4}) holds because, when $a \notin \mathcal{A}_{ij}$, we have $\Delta Z_{ij}(n) \equiv 0$. The penultimate inequality is a consequence of the fact that, under $\pi_{SA}(L,\eta)$, one will choose an action $a \in \mathcal{A}_{ij}$ with probability at least $\eta \beta$.
\end{IEEEproof}

We now proceed to show  an inequality analogous to the Chernoff bound for the log-likelihood ratio. In classical sequential hypothesis testing, due to independence of samples across time, the expectation of the likelihood ratio can be split as the product of the expectation of the likelihood ratio increments, as follows:
\begin{align*}
 E_{i} \left[e^{sZ_{ji}(n)} \right] = \prod_{k=1}^{n} E_{i} \left[e^{s\Delta Z_{ji}(n)} \right].
\end{align*}
The same decomposition is not valid in ASHT because actions introduce dependency in the likelihood ratio increments across time. However, we can obtain an upper bound of the product form.

\begin{lemma}
\label{lemma:likelihood ratio decomposition}
 Assume (I) and (IIb). Consider policy $\tilde{\pi}_{SA}(\eta)$. Fix $i$. We then have, for all $0 <s<1$,
\begin{align*}
 E_{i} \left[e^{s Z_{ji}(n)}\right] \le (\rho_{ij}(s))^{n} \; \; \; \hfill \forall n, \forall j \ne i.
\end{align*}
\end{lemma}

\begin{IEEEproof}
Once again, we proceed through the chain of inequalities all of which are now self-evident:
\begin{align*}
 E_{i} &\left[e^{s Z_{ji}(n)} \right]\\
&= E_{i} \left[ E_{i} \left[ e^{s Z_{ji}(n-1)}e^{s\Delta Z_{ji}(n)} \vert X^{n-1},A^{n-1} \right] \right]\\
&=  E_{i} \left[e^{s Z_{ji}(n-1)} E_{i} \left[e^{s\Delta Z_{ji}(n)}\vert X^{n-1},A^{n-1} \right]  \right] \\
& = \rho_{ij}(s) E_{i} \left[e^{s Z_{ji}(n-1)}\right] \text{\hfill (from Lemma \ref{lemma:likelihood ratio property under policy})}\\
&\le (\rho_{ij}(s))^{n},
\end{align*}
where the last inequality follows by induction.
\end{IEEEproof}

We now show an exponential decay property of the log-likelihood process which primarily stems from the anticipated negative drift in $Z_{ji}(n)$ for $j\ne i$. Let us alert the reader that in the following Proposition we deal with $Z_{ij}(n) = -Z_{ji}(n)$.

\begin{proposition}
\label{prop:exponential decay LLR process}
 Assume (I) and (IIb). Consider policy $\tilde{\pi}_{SA}(\eta)$. Fix $i$. There exist constants $C_{K} > 0$ and $\gamma > 0$ such that
\begin{align}
\label{eqn:exp decay LLR ratio}
 P_{i}\left(\min_{j\ne i} Z_{ij}(n) \le K \right) < C_{K} e^{-\gamma n}.
\end{align}
$C_{K}$ is independent of $i$, but $\gamma$ may depend on $i$.
\end{proposition}

\begin{IEEEproof}
This follows from the previous lemmas via the following :
 \begin{align}
 \nonumber P_{i}\left(\min_{j\ne i} Z_{ij}(n) \le K \right) &= P_{i}\left(\max_{j\ne i} Z_{ji}(n) \ge -K \right)\\
 \label{eqn: exponential decay LLR process inequality 3} &\le \sum_{j \ne i} P_{i}\left(Z_{ji}(n) \ge -K \right)\\
 \label{eqn: exponential decay LLR process inequality 1} & \le \sum_{j \ne i} e^{sK}E_{i}\left[e^{sZ_{ji}(n)}\right]\\
 \label{eqn: exponential decay LLR process inequality 2} & \le e^{sK} \sum_{j \ne i} (\rho_{ij}(s))^{n}\\
 \nonumber & \le e^{sK} \cdot (M-1) \cdot \max_{j\ne i} (\rho_{ij}(s))^{n}\\
 \nonumber & = C_{K} e^{-\gamma n},
 \end{align}
where $\max_{j \ne i} \rho_{ij}(s) = e^{-\gamma}$, and $C_{K} = M e^{sK}$. The inequality in (\ref{eqn: exponential decay LLR process inequality 3}) is due to the union bound, the inequality in (\ref{eqn: exponential decay LLR process inequality 1}) is due to Chernoff's bound with $0<s<1$,  and the inequality in (\ref{eqn: exponential decay LLR process inequality 2}) is due to Lemma \ref{lemma:likelihood ratio decomposition}.
\end{IEEEproof}
%%%%%%%%%%%%%%%%%%%%%%%%%%%%%%%%%%%%%%%%%%%%%%%%%%%%%%%%%%%%%%%%%%%%%%%%%%%%%%%%%%%%

%%%%%%%%%%%%%%%%%%%%%%%%%%%%%%%%%%%%%%%%%%%%%%%%%%%%%%%%%%%%%%%%%%%%%%%%%%%%%%%%%%%%

 We now show that under the hypothesis $H = H_{i}$, the $\theta(n)$ process eventually settles at $i$.  Indeed we show something stronger. Let us define
\begin{align}
\label{eqn:Ti}
 T_{i} := \inf\{n:\theta(n') = i, \quad \forall n' \ge n\},
\end{align}
the time at which $\theta(n)$  meets its eventuality of settlement at $i$. This random variable has a tail that decays exponentially fast, as shown next.

\begin{lemma}
\label{lemma:exponential decay of Ti}
 Assume (I) and (IIb). Consider policy $\tilde{\pi}_{SA}(\eta)$. Fix $i$. Then there exist $C>0$ and $b >0$, both finite and possibly dependent on $i$, such that
\begin{align}
\label{eqn:exp decay Ti}
 P_{i}\left(T_{i} > n \right) < C e^{-bn}. \quad
\end{align}
\end{lemma}

\begin{IEEEproof}
By the union bound
 \begin{align*}
 P_{i}\left(T_{i} > n \right) & = P_{i}(\theta(n') \ne i \text{ for some $n' \ge n$})\\
 & \le  \sum_{n' \ge n}P_{i}\left(\theta(n') \ne i\right)\\
 & \le \sum_{n' \ge n}P_{i}\left(\min_{j \ne i} Z_{ij}(n') \le 0\right).
 \end{align*}
The assertion now follows from Proposition \ref{prop:exponential decay LLR process}.
\end{IEEEproof}

%%%%%%%%%%%%%%%%%%%%%%%%%%%%%%%%%%%%%%%%%%%%%%%%%%%%%%%%%%%%%%%%%%%%%%%%%%%%%%%%%%%%

%%%%%%%%%%%%%%%%%%%%%%%%%%%%%%%%%%%%%%%%%%%%%%%%%%%%%%%%%%%%%%%%%%%%%%%%%%%%%%%%%%%%
Thus far we have considered the policy $\tilde{\pi}_{SA}(\eta)$ which never stops. We now show that the policy $\pi_{SA}(L,\eta)$ stops in finite time.
\begin{proposition}
\label{prop:finite stopping time}
 Assume (I) and (IIb). Consider the policy $\pi_{SA}(L,\eta)$. Fix $i$. We then have
\begin{align*}
 P_{i}(\tau(\pi_{SA}(L,\eta)) < \infty) = 1.
\end{align*}
\end{proposition}
\begin{IEEEproof}
 We consider $\pi_{SA}^{i}(L,\eta)$  for analysis. Recall that $\tau(\pi_{SA}(L,\eta)) \le \tau(\pi_{SA}^{i}(L,\eta))$, and hence it is sufficient to show that
\begin{align}
\label{eqn:stopping time pi_i finite}
 P_{i}(\tau(\pi_{SA}^{i}(L,\eta) < \infty) = 1.
\end{align}
From Proposition \ref{prop:exponential decay LLR process}, we know that, for a suitable constant $\tilde{C}$, $$P_{i}\left(\min_{j\ne i} Z_{ij}(n) < \log(L(M-1)) \right) < \tilde{C} e^{-\gamma n}.$$ Since this bound is summable, by the Borel-Cantelli lemma, $$P_{i}\left(\min_{j\ne i} Z_{ij}(n) < \log(L(M-1)) \quad \text{infinitely often}\right) = 0,$$ which is stronger than the assertion (\ref{eqn:stopping time pi_i finite}).
\end{IEEEproof}
Propositions \ref{prop:exponential decay LLR process} and \ref{prop:finite stopping time} are the ones that will be used in the sequel.
%%%%%%%%%%%%%%%%%%%%%%%%%%%%%%%%%%%%%%%%%%%%%%%%%%%%%%%%%%%%%%%%%%%%%%%%%%%%%%%%%%%%

%%%%%%%%%%%%%%%%%%%%%%%%%%%%%%%%%%%%%%%%%%%%%%%%%%%%%%%%%%%%%%%%%%%%%%%%%%%%%%%%%%
\subsection{Proof of Proposition \ref{prop:probability of wrong detection}}
\label{appendix:proof of probability of wring detection}

The proof relies on a standard change of measure argument. Let $\Delta_{j}$ denote the event that the policy $\pi_{SA}(L,\eta)$ declares $H_{j}$ as the true hypothesis.
 \begin{align}
\nonumber  P_{i}(\delta \ne i) &= \sum_{j \ne i} P_{i}(\delta = j)+ P_{i}(\tau(\pi_{SA}(L,\eta)) = \infty)\\
\nonumber &= \sum_{j \ne i} \sum_{n>0}\int_{\omega^{n} \in \Delta_{j}} dP_{i}(\omega^{n}) + 0\\
\nonumber &= \sum_{j \ne i} \sum_{n>0}\int_{\omega^{n} \in \Delta_{j}} \frac{dP_{i}}{dP_{j}}(\omega^{n})dP_{j}(\omega^{n})\\
\label{eqn:proof of propostion prob of wrong detection equation 4} &\le \sum_{j \ne i} \sum_{n>0}\int_{\omega^{n} \in \Delta_{j}} \frac{1}{(M-1)L}dP_{j}(\omega^{n})\\
\nonumber& \le \frac{1}{(M-1)L} \sum_{j \ne i} P_{j}(\Delta_{j})\\
\nonumber & \le \frac{1}{L}.
 \end{align}
The second equality holds because we have shown in Proposition \ref{prop:finite stopping time} that the stopping time is finite with probability 1. The inequality (\ref{eqn:proof of propostion prob of wrong detection equation 4}) follows because $\omega^{n} \in \Delta_{j}$ implies $Z_{ji}(n)(\omega^{n}) \ge \log((M-1)L)$, that is, $\frac{dP_{i}}{dP_{j}}(\omega^{n}) \le \frac{1}{(M-1)L}$.
\hfill \IEEEQEDclosed
%%%%%%%%%%%%%%%%%%%%%%%%%%%%%%%%%%%%%%%%%%%%%%%%%%%%%%%%%%%%%%%%%%%%%%%%%%%%%%%%%%

%%%%%%%%%%%%%%%%%%%%%%%%%%%%%%%%%%%%%%%%%%%%%%%%%%%%%%%%%%%%%%%%%%%%%%%%%%%%%%%%%%%%%%%%%%
\subsection{Proof of Theorem \ref{theorem: upper bound on stopping time of SA}: Achievability}
\label{ref:lemma:achievability proof}
We assume (I) and (IIb). All statements in this proof are under $H = H_{i}$ and under {\it Sluggish Procedure A}.  We follow the proof technique of Chernoff \cite[Lem. 2]{ref:195909AMS_Che}. Chernoff's proof technique does not go through completely because unlike in {\it Procedure A}, the next action in {\it Sluggish Procedure A} is not conditionally independent of the previous action, given the current likelihood values. A similar issue was addressed by Nitinawarat and Veeravalli in \cite{2013arXiv1310.1844_NitinawaratVeeravalli}, in the context of Markovian observation model, and we will adapt their proof technique to our setting.

Let us first setup some notation. Fix $\epsilon > 0$. Define $$D_{ij}:= \sum_{a \in \mathcal{A}} \lambda_{i}(a) D(q_{i}^{a} \Vert q_{j}^{a}),$$ where $\lambda_{i}$ is as defined in (\ref{eqn:optimal_lambda}). Let $D_{i}$ be as defined by (\ref{eqn:D_i}), i.e., $D_{i} = \min_{j \ne i}D_{ij}$. Under the {\it Sluggish Procedure A}, the transition probability matrix $TP(\theta({n}))$ of the action process $A_{n}$ at time $n$ is given by

\begin{align}
\label{eqn:transition_prob_matrix}
 TP(\theta(n)) = (1-\eta) \mathbf{I}+\eta \left(\mathbf{\underbar{1}} \: \lambda_{\theta(n)}^{T}\right).
\end{align}

It is easy to verify that the stationary distribution associated with $TP(\theta({n}))$ is $\lambda_{\theta(n)}$. Define $\mathcal{F}_{k-1} := \sigma(X^{k-1},A^{k-1})$, the $\sigma$-field generated by the random variables $(X^{k-1},A^{k-1})$.

We now upper bound the expected time to make a decision under {\it Sluggish Procedure A} as follows:
\begin{align}
 \nonumber E_{i}\left[\tau(\pi_{SA}(L,\eta))\right] & \le E_{i}\left[\tau(\pi_{SA}^{i}(L,\eta))\right]\\
 \nonumber & = \sum_{n \ge 0} P_{i} \left(\tau(\pi_{SA}^{i}(L,\eta)) > n\right)\\
  \nonumber & \le \frac{(1+\epsilon) \log (L(M-1))}{D_{i}}\\
\label{eqn:bounding expected stopping time equation 3} & \hspace{0.5 cm}+ \sum_{n \ge \tilde{n}} P_{i} \left(\tau(\pi_{SA}^{i}(L,\eta)) > n\right),
 \end{align}
where $$\tilde{n} =  \frac{(1+\epsilon) \log (L(M-1))}{D_{i}}.$$
To complete the proof, we will now show that for any $\epsilon >0$, the second term on the right-hand side of (\ref{eqn:bounding expected stopping time equation 3}) goes to zero as $L \rightarrow \infty$. Indeed, we claim that each term in the summation decays exponentially with $n$ with an exponent that does not depend on $L$. Assuming the claim, the tail sum vanishes as $L \rightarrow \infty$, because $\tilde{n} \rightarrow \infty$. This suffices to complete the proof of Theorem \ref{theorem: upper bound on stopping time of SA}.

We now proceed to prove the claim. Observe that
\begin{align*}
 P_{i} &\left(\tau(\pi_{SA}^{i}(L,\eta)) > n\right)\\
 & \le P_{i} \left(\min_{j \ne i} Z_{ij}(n) \le \log(L(M-1))\right)\\
& \le \sum_{j \ne i}  P_{i} \left(Z_{ij}(n) \le \log(L(M-1))\right).
\end{align*}

Fix one $j \ne i$. (The same analysis holds for other $j$.) Then
{
\begin{align}
 \nonumber P_{i}  & \left(Z_{ij}(n) \le \log(L(M-1))\right)\\
 \nonumber & = P_{i} \left(\sum_{k=1}^{n}\Delta Z_{ij}(k) \le \log(L(M-1))\right) \\
  \nonumber& = P_{i} \left(\sum_{k=1}^{n} \left( \Delta Z_{ij}(k) - E_{i} \left[ \Delta Z_{ij}(k)\vert \mathcal{F}_{k-1}\right] + \epsilon' \right) \right. \\
  \nonumber & \hspace{1.2 cm} + \sum_{k=1}^{n} \left(E_{i} \left[ \Delta Z_{ij}(k)\vert \mathcal{F}_{k-1}\right] - D_{ij} + \epsilon' \right) \\
  \nonumber & \hspace{1.2 cm} +  n \left(D_{ij} - 2\epsilon' \right) \le \log (M-1)L \Bigg) \\
  \nonumber & \le P_{i}\left( \sum_{k=1}^{n} \left( \Delta Z_{ij}(k) - E_{i} \left[ \Delta Z_{ij}(k)\vert \mathcal{F}_{k-1}\right] + \epsilon' \right) < 0 \right) \\
  \nonumber& \hspace{0.7 cm} + P_{i} \left( \sum_{k=1}^{n} \left(E_{i} \left[ \Delta Z_{ij}(k)\vert \mathcal{F}_{k-1}\right] - D_{ij} + \epsilon' \right) < 0 \right) \\
 \label{eqn:log likelihood ratio Z_ij equation 2} &  \hspace{0.7 cm} + P_{i} \left( n (D_{ij} - 2\epsilon') \le \log(L(M-1))  \right).
\end{align}
}Look at the first probability term in (\ref{eqn:log likelihood ratio Z_ij equation 2}). Each entry within the summation has a positive mean and, from Chernoff's bounding technique in \cite[Lem. 2]{ref:195909AMS_Che}, there exists a $b(\epsilon') > 0$ such that
\begin{align*}
 P_{i} & \left( \sum_{k=1}^{n} \left( \Delta Z_{ij}(k) - E_{i} \left[ \Delta Z_{ij}(k)\vert \mathcal{F}_{k-1}\right] + \epsilon' \right) < 0 \right) \le e^{-n b(\epsilon')}.
\end{align*}

The third probability term is $0$ if we choose an $\epsilon'$ small enough such that $n(D_{ij} - 2 \epsilon') > \log (L(M-1))$, for all $n > \tilde{n}$. Indeed, any $\epsilon'$ satisfying $0 < \epsilon' < \frac{\epsilon}{1+\epsilon}\frac{D_{i}}{2}$ suffices. So set $\epsilon' = \frac{\epsilon}{1+\epsilon} \frac{D_{i}}{4}$.

We now proceed to show that the second term also decays exponentially to zero. Let $T_{i}$ be as defined in (\ref{eqn:Ti}). For a suitably chosen $\epsilon''$, and we will soon indicate how to choose it, we have
{\small
\begin{align*}
 P_{i} & \left( \sum_{k=1}^{n} \left(E_{i} \left[ \Delta Z_{ij}(k)\vert \mathcal{F}_{k-1}\right] - D_{ij} + \epsilon' \right) < 0 \right)  \\
& \le P_{i} \left( \sum_{k=1}^{n} \left(E_{i} \left[ \Delta Z_{ij}(k)\vert \mathcal{F}_{k-1}\right] - D_{ij} + \epsilon' \right) < 0,~ T_{i} \le n\epsilon'' \right)\\
& \hspace{0.7 cm} + P_{i}(T_{i} > n \epsilon'').
\end{align*}
}
From Lemma \ref{lemma:exponential decay of Ti}, the second probability term on the right-hand side decays exponentially with $n$. To show that the first probability term on the right-hand side decays exponentially with $n$, we use a technique of Nitinawarat and Veeravalli \cite[(6.23)]{2013arXiv1310.1844_NitinawaratVeeravalli}.

First, we indicate how to choose $\epsilon''$. Define
\begin{align*}
\tilde{C} & = \min_{a \in \mathcal{A}} E_{i}\left[\Delta Z_{ij}(k)\vert A_{k}=a \right] - D_{ij}\\
& = \min_{a \in \mathcal{A}} D(q_{i}^{a}\Vert q_{j}^{a}) - D_{ij}.
\end{align*}
Since $D_{ij}$ is the $\lambda_{i}$-weighted average of $D(q_{i}^{a}\Vert q_{j}^{a})$, we have $\tilde{C} \le 0$. Choose $\epsilon''$ small enough so that $\tilde{\epsilon} := \epsilon'+\epsilon''\tilde{C} > 0$. We then have
{
\begin{align*}
 P_{i}  & \left( \sum_{k=1}^{n} \left(E_{i} \left[ \Delta Z_{ij}(k)\vert \mathcal{F}_{k-1}\right] - D_{ij} + \epsilon' \right) < 0,~ T_{i} \le n\epsilon'' \right) \\
 & = P_{i} \left(\sum_{k=1}^{\lfloor n \epsilon'' \rfloor} \left(E_{i} \left[ \Delta Z_{ij}(k)\vert \mathcal{F}_{k-1}\right] - D_{ij} + \epsilon' \right) \right.\\
 & \hspace{1.2 cm}+ \sum_{ k= \lfloor n \epsilon'' \rfloor + 1}^{n} \left(E_{i} \left[ \Delta Z_{ij}(k)\vert \mathcal{F}_{k-1}\right] - D_{ij} + \epsilon' \right)  < 0,\\
 &  \hspace{1.2 cm} T_{i} \le n\epsilon'' \Bigg)\\
 & \le P_{i} \left(\lfloor n \epsilon'' \rfloor (\tilde{C}+\epsilon') \right.\\
 &  \hspace{1.2 cm} + \sum_{k= \lfloor n \epsilon'' \rfloor + 1}^{n} \left(E_{i} \left[ \Delta Z_{ij}(k)\vert \mathcal{F}_{k-1}\right] - D_{ij} + \epsilon' \right)  < 0, \\
 & \hspace{1.2 cm} T_{i} \le n\epsilon'' \Bigg)\\
 & \le P_{i} \left( \sum_{k= \lfloor n \epsilon'' \rfloor + 1}^{n} \left(E_{i} \left[ \Delta Z_{ij}(k)\vert \mathcal{F}_{k-1}\right] - D_{ij} +  \tilde{\epsilon} \right)  < 0, \right. \\
 &  \hspace{1.2 cm} T_{i} \le n\epsilon'' \Bigg)\\
 & \le \tilde{P}_{i} \left( \sum_{k=\lfloor n \epsilon'' \rfloor + 1}^{n} \left(E_{i} \left[ \Delta Z_{ij}(k)\vert \mathcal{F}_{k-1}\right] - D_{ij}+ \tilde{\epsilon} \right) < 0 \right)\\
 & \le C e^{-n\tilde{b}(\tilde{\epsilon})}, \numberthis \label{eqn:exponential decay stationary markov process equation 5}
\end{align*}
}for some $C > 0$ and some $\tilde{b}(\tilde{\epsilon}) > 0$. The second inequality follows from the fact that $\tilde{C} \le E_{i} \left[ \Delta Z_{ij}(k)\vert \mathcal{F}_{k-1}\right] - D_{ij}$, for all $k$. The third inequality follows from the choice of $\tilde{\epsilon}$ and the fact that $$\lfloor n \epsilon'' \rfloor (\tilde{C}+\epsilon')+ (n - \lfloor n \epsilon'' \rfloor) \epsilon' \ge (n-\lfloor n \epsilon'' \rfloor) \tilde{\epsilon}.$$ $\tilde{P}_{i}$ is a new measure under which actions are taken according to {\it Sluggish Procedure A} but assuming $\theta(n) = i \quad \forall n$, and the observations are conditionally independent of past observations and actions, given the current action. Consequently, under $\tilde{P}_{i}$, the action process $A_{n}$ is a stationary Markov Chain with transition probability matrix $TP(i)$. By the ergodic theorem and concentration inequalities for Markov Chains \cite{1998_AnnalsAppliedProb_Lezaud}, this term also decays exponentially with $n$, which is (\ref{eqn:exponential decay stationary markov process equation 5}).
\hfill \IEEEQEDclosed

%%%%%%%%%%%%%%%%%%%%%%%%%%%%%%%%%%%%%%%%%%%%%%%%%%%%%%%%%%%%%%%%%%%%%%%%%%%%%%%%%%%%%%%%%%

%%%%%%%%%%%%%%%%%%%%%%%%%%%%%%%%%%%%%%%%%%%%%%%%%%%%%%%%%%%%%%%%%%%%%%%%%%%%%%%%%%%%%%%%%%
%%%%%%%%%%%%%%%%%%%%%%%%%%%%%%%%%%%%%%%%%%%%%%%%%%%%%%%%%%%%%%%%%%%%%%%%%%%%%%%%%%%%%%%%%%%%%
%%%%%%%%%%%%%%%%%%%%%%%%%%%%%%%%%%%%%%%%%%%%%%%%%%%%%%%%%%%%%%%%%%%%%%%%%%%%%%%%%%%%%%%%%%
\section{Proof of Proposition \ref{prop:optimum Di for Case 2}}
\label{proof:proof of optimum Di}
We will focus only on $i \le W$ and will determine
\begin{align}
 \label{eqn:optimum Di}
 D_{i} &= \max_{\lambda \in \mathcal{P}(\mathcal{A})} \min_{j \ne i} \sum_{a \in \mathcal{A}} \lambda(a) D(q_{i}^{a} \Vert q_{j}^{a}).
\end{align}
The case $i > W$ can be handled similarly and is omitted. Using (\ref{eqn:KL1}) - (\ref{eqn:KL3}),  we can simplify the minimisation in (\ref{eqn:optimum Di}) by considering three regions for $j$ as follows:
\begin{align}
\nonumber D_{i} &= \max_{\lambda \in \mathcal{P}(\mathcal{A})} \min \left\{ \min_{j \le W , j \ne i} \left(\lambda({i}) D(f_{k} \Vert f_{l})+ \lambda({j}) D(f_{l} \Vert f_{k}) \right), \right.\\
\nonumber & \hspace{3 cm} \lambda({i}) D(f_{k} \Vert f_{l})+ (1-\lambda({i})) D(f_{l} \Vert f_{k}),\\
\label{eqn:optimum D simplified 1a}& \hspace{3 cm} \left. \min_{j > W , j \ne i+W} (1-\lambda({i})-\lambda({j-W})) D(f_{l} \Vert f_{k}) \right\},\\
\nonumber &= \max_{\lambda \in \mathcal{P}(\mathcal{A})} \min \left\{\lambda({i}) D(f_{k} \Vert f_{l})+ \min_{ j \ne i} \lambda({j}) D(f_{l} \Vert f_{k}), \right.\\
\nonumber & \hspace{3 cm} \lambda({i}) D(f_{k} \Vert f_{l})+ (1-\lambda({i})) D(f_{l} \Vert f_{k}),\\
\label{eqn:optimum D simplified 1b}& \hspace{3 cm} \left. (1-\lambda({i})-\max_{j \ne i} \lambda({j})) D(f_{l} \Vert f_{k}) \right\}.
\end{align}
Observe that the second term is always greater than or equal to the other two terms, and hence can be removed from the minimisation. Thus,
\begin{align}
\nonumber D_{i} \nonumber &= \max_{\lambda \in \mathcal{P}(\mathcal{A})} \min \left\{\lambda({i}) D(f_{k} \Vert f_{l})+ \min_{ j \ne i} \lambda({j}) D(f_{l} \Vert f_{k}), \right.\\
\label{eqn:optimum D simplified 2}& \hspace{3 cm} \left. (1-\lambda({i})-\max_{j \ne i} \lambda({j})) D(f_{l} \Vert f_{k}) \right\}.
\end{align}
We now perform the maximisation over $\lambda$ in two steps. First, let us fix $\lambda (i)$ and optimise over the distribution of $1-\lambda (i)$ among the other actions. Since $$\min_{j \ne i} \lambda (j) \le \frac{1-\lambda (i)}{W-1} \le \max_{j \ne i} \lambda (j),$$ we have
$$\left(\min_{j \ne i} \lambda (j)\right) D(f_{l} \Vert f_{k}) \le \left(\frac{1-\lambda (i)}{W-1}\right) D(f_{l}\Vert f_{k})$$ and $$-\max_{j \ne i} \lambda (j) D(f_{l} \Vert f_{k}) \le -\frac{(1-\lambda (i))}{W-1} D(f_{l} \Vert f_{k}).$$
Thus both the terms within braces in (\ref{eqn:optimum D simplified 2}) are lesser than or equal to the corresponding terms for equal distribution of $1-\lambda (i)$ among the other actions. The optimisation problem is now reduced to a single variable optimisation of the form
\begin{align}
\nonumber D_{i} \nonumber &= \max_{0\le \lambda(i) \le 1} \min \left\{\lambda({i}) D(f_{k} \Vert f_{l})+ \frac{(1- \lambda({i}))}{W-1} D(f_{l} \Vert f_{k}), \right.\\
\label{eqn:optimum D simplified 3}& \hspace{3 cm} \left. (1-\lambda({i}))\frac{W-2}{W-1} D(f_{l} \Vert f_{k}) \right\}.
\end{align}
Second, we now perform the optimisation in (\ref{eqn:optimum D simplified 3}) over $\lambda(i)$. The first term in the minimisation is increasing or non-increasing in $\lambda(i)$ depending on $D(f_{k}\Vert f_{l}) > D(f_{l}\Vert f_{k})/(W-1)$ or $D(f_{k}\Vert f_{l}) \le {D(f_{l}\Vert f_{k})}/{(W-1)}$, respectively. The second term is decreasing in $\lambda(i)$.

\textit{1)} Suppose  $D(f_{k}\Vert f_{l}) > {D(f_{l}\Vert f_{k})}/{(W-1)}$, then the two terms viewed as linear functions over $\lambda(i)$ cross each other, and so the maximum will be achieved at the point of equality, i.e.,
\begin{align*}
\lambda({i}) &D(f_{k} \Vert f_{l})+ \frac{(1- \lambda({i}))}{W-1}  D(f_{l} \Vert f_{k}) \\ =  &(1-\lambda({i}))\frac{W-2}{W-1} D(f_{l} \Vert f_{k}).
\end{align*}
Solving for $\lambda(i)$ yields
\begin{align*}
 \lambda(i)&=\frac{(W-3)D(f_{l}\Vert f_{k})}{(W-1) D(f_{k}\Vert f_{l})+(W-3) D(f_{l} \Vert f_{k})},\\
\lambda(j)&=\frac{D(f_{k}\Vert f_{l})}{(W-1) D(f_{k}\Vert f_{l})+(W-3) D(f_{l} \Vert f_{k})}, ~ \forall ~ j \ne i,\\
D_{i} &= \frac{(W-2)D(f_{k}\Vert f_{l})D(f_{l}\Vert f_{k})}{(W-1) D(f_{k}\Vert f_{l})+(W-3) D(f_{l} \Vert f_{k})}.
\end{align*}

\textit{2)} Suppose $D(f_{k}\Vert f_{l}) \le {D(f_{l}\Vert f_{k})}/{(W-1)}$, then the maximum is achieved at $\lambda(i) =0$. Then $\lambda (j) = {1}/{(W-1)}, ~ \forall j \ne i$, and
\begin{align*}
 D_{i} &= \min \left\{\frac{D(f_{l} \Vert f_{k})}{W-1},\frac{(W-2)D(f_{l} \Vert f_{k})}{W-1} \right\}\\
& = \frac{D(f_{l} \Vert f_{k})}{W-1},
\end{align*}
since $W > 3$.
\IEEEQEDclosed

\section{Estimation of Relative Entropy Rate}
\label{sec:app:EstimateRelEntropy}

The computation of our proposed neuronal index requires a computation of the relative entropy rate between two Poisson point processes from estimates of their rates. The relative entropy between two Poisson point processes with rates $R_{1}$ and $R_{2}$ is 
\begin{align}
\label{eqn:KL distance poisson}
 \nonumber \frac{1}{T} D(\mu_{R_{1},T}\Vert \mu_{R_{2},T}) &= R_{1} \log\left(\frac{R_{1}}{R_{2}}\right)+R_{2}-R_{1}\\
 &=  R_{1} \log R_{1}-R_{1} \log R_{2}+R_{2}-R_{1}.
\end{align}
Let $N_{i}(k,T)$ be the number of spikes observed in time slot $k$, $1\le k \le n$, of duration $T$ on the $i^{th}$ process, $i = 1, 2$. The empirical firing rate is then $\hat{R}_{i} = \frac{1}{nT}\sum_{k=1}^{n} N_{i}(k,T)$. A natural estimate for (\ref{eqn:KL distance poisson}), based on the observations, would be to substitute $R_{i}, ~ i = 1,2$ by their respective empirical estimates $\hat{R}_{i}, ~ i= 1,2$, to get

\begin{align}
 \nonumber \hat{D} & = \hat{R}_{1} \log \left(\frac{\hat{R}_{1}}{\hat{R}_{2}}\right)+\hat{R}_{2}-\hat{R}_{1}\\
    \label{eqn:relative entropy plug-and-play estimator} & = \hat{R}_{1} \log\hat{R}_{1} - \hat{R}_{1} \log\hat{R}_{2}+\hat{R}_{2}-\hat{R}_{1}.
\end{align}
A little reflection suggests that this is a bad estimate, for there is a positive probability that $\hat{R}_{1} > \hat{R}_{2} = 0$, yielding  $E_{R_{1},R_{2}} \left[ \hat{D}\right] = \infty $. Estimate (\ref{eqn:relative entropy plug-and-play estimator}) is thus biased (though consistent). Our approach is to obtain estimates for each of the terms in (\ref{eqn:KL distance poisson}) with minimal bias.

\begin{figure}[t]
 \centering
\includegraphics[scale=0.6]{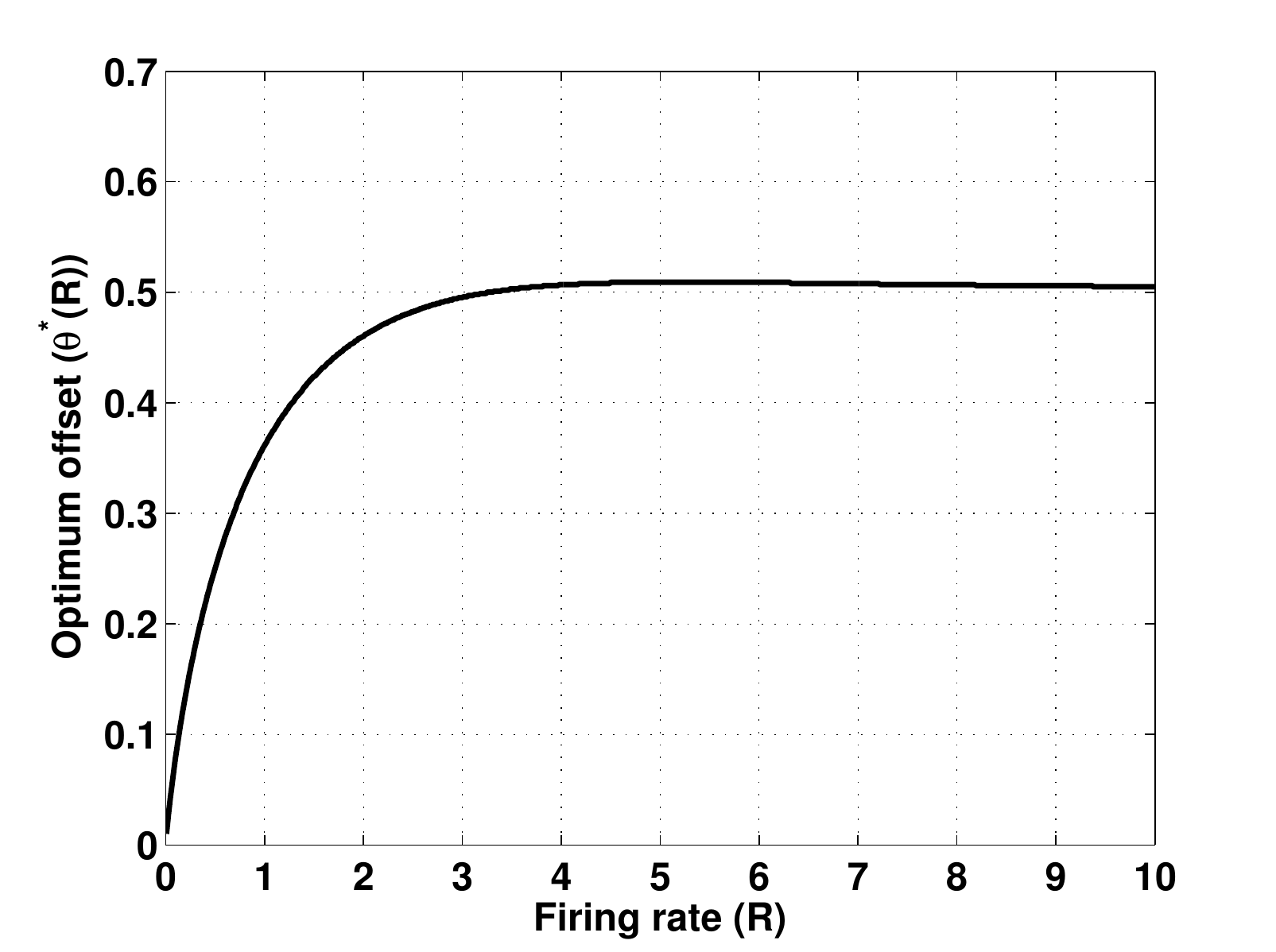}
\caption{Optimum offset ($\theta^{*}(R)$) to minimise bias for different firing rates (R).}
\label{fig:optimum_theta_different_rates}
\end{figure}

Unbiased and maximum likelihood estimates for the third and fourth terms on the right hand side of (\ref{eqn:KL distance poisson}) are the respective empirical firing rates themselves. Let us therefore now study the second term. We may assume that the firings are independent, given $R_1$ and $R_2$. Thus we may look for an estimator of the form $-\hat{R}_1 f(\hat{R}_2)$ which has expectation $-R_{1} E_{R_2}[f (\hat{R}_{2})]$. For this to be close to the desired $-R_1 \log R_2$, we look for a function $f(\hat{R}_{2})$ such that $E_{R_{2}}\left[f(\hat{R}_{2})\right] \approx \log R_{2}$. The difficulty is due to the $\log(0) = -\infty$ artifact. We consider a simple fix of adding a nonzero offset to the empirical estimate, i.e., we consider estimates of the form $\log(\hat{R}+ \theta)$. Figure \ref{fig:optimum_theta_different_rates} shows the optimum offset $\theta^{*}(R)$ for different firing rates $R$ when $n=T=1$. The optimum offset $\theta^{*}(R)$ can be seen to converge to 0.5 for large $R$. Further, the convergence is quite fast, $\theta^{*}(R)$ is close to 0.5 for all $R$ greater than 3.  Hence in this work we use $\theta = 0.5$ as the offset, thus resulting in an estimator for $\log(R)$ of the form $\log(\hat{R}+1/2)$. {For a general $n$ and $T$ we then have $E\left[\log(nT\hat{R}+1/2)\right] \approx \log(nTR)$, which in turn implies $E\left[\log(\hat{R} + 1/2nT)\right] \approx \log(R)$. Thus an estimator for a general $n$ and $T$ would be $\log\left(\hat{R}+{1}/{2nT}\right)$. The estimator for the second term in (\ref{eqn:KL distance poisson}) is then $-\hat{R}_{1} \log \left(\hat{R}_{2}+{1}/{2nT}\right)$. One could look for better estimators with the offset being a function of the observed empirical means. In this work we stick to the constant offset estimator, with the constant offset being $\theta = 0.5$, as it is reasonable to assume that the neurons have a firing rate greater than $3/nT = 3/(24*0.25) = 0.5$ spikes/second ($n$ = 24, $T$ = 250 ms), thus putting them in the firing rate regime where $\theta = 0.5$ is a good offset for near unbiasedness. The values $T$ = 250ms and $n$ = 24 correspond to the neuronal recording time and the number of repetitions in the neuronal recording experiment of Sripati and Olson \cite{ref:201001JNS_SriOls}.} 

To address the first term of (\ref{eqn:KL distance poisson}) we consider estimates of the form $\hat{R}_{1}g(\hat{R}_{1})$ such that
\begin{align*}
 E_{R_{1}}\left[\hat{R}_{1} g(\hat{R}_{1})\right] \cong R_{1} \log R_{1}.
\end{align*}
Expanding the expectation above for $n=T=1$, we obtain,
\begin{align*}
 E_{R_{1}}\left[\hat{R}_{1} g(\hat{R}_{1})\right] &= \sum_{k=0}^{\infty} k g(k) \frac{{R_{1}}^{k} e^{-R_{1}}}{k!}\\
 &= R_{1} \sum_{k=1}^{\infty} g(k)\frac{R_{1}^{k-1} e^{-R_{1}}}{(k-1)!}\\
 &= R_{1} E_{R_{1}}\left[g(\hat{R}_{1}+1)\right].
\end{align*}
Thus we want a $g$ such that $E_{R_{1}}\left[g(\hat{R}_{1}+1)\right] \cong \log R_{1}$. From the discussion on the second term, we know that $E_{R_{1}}\left[\log(\hat{R}_{1}+\frac{1}{2})\right] \sim \log R_{1}$, and hence a good choice for $g$ would be $g(\hat{R}_{1}) = \log(\hat{R}_{1}-\frac{1}{2})$. Thus our estimate for the first term for a general $n$ and $T$ is
\begin{align*}
\begin{cases}
  \hat{R}_{1} \log\left(\hat{R}_{1}-\frac{1}{2nT}\right) &\text{if } \hat{R}_{1} \ge \frac{1}{2nT} \text{, i.e., there is atleast one point},\\
  0 & \text{otherwise}.
\end{cases}
\end{align*}

Therefore our combined estimate for the relative entropy rate in (\ref{eqn:KL distance poisson}), based on the average firing rate estimates $\hat{R}_{1}$ and $\hat{R}_{2}$ and obtained over a time of duration $nT$ is
\begin{align}
 \nonumber \hat{D}&(\hat{R}_{1}\Vert \hat{R}_{2}) =
 \begin{cases}
  \left[\hat{R}_{1}\log\left(\frac{\hat{R}_{1}-\frac{1}{2nT}}{\hat{R}_{2}+\frac{1}{2nT}}\right)+\hat{R}_{2}-\hat{R}_{1}\right]^{+} &\text{if } \hat{R}_{1} \ge \frac{1}{2nT}\\
   \hat{R_{2}} & \text{otherwise}.
 \end{cases}
\end{align}
Relative entropy being a convex function of its arguments, the plug-in estimator of (\ref{eqn:relative entropy plug-and-play estimator}) would always have a positive bias. Naturally, an unbiased estimator would have a smaller value than (\ref{eqn:relative entropy plug-and-play estimator}), and our proposed estimator does satisfy this requirement.

\begin{figure}[ht]
 \centering
\includegraphics[scale=0.6]{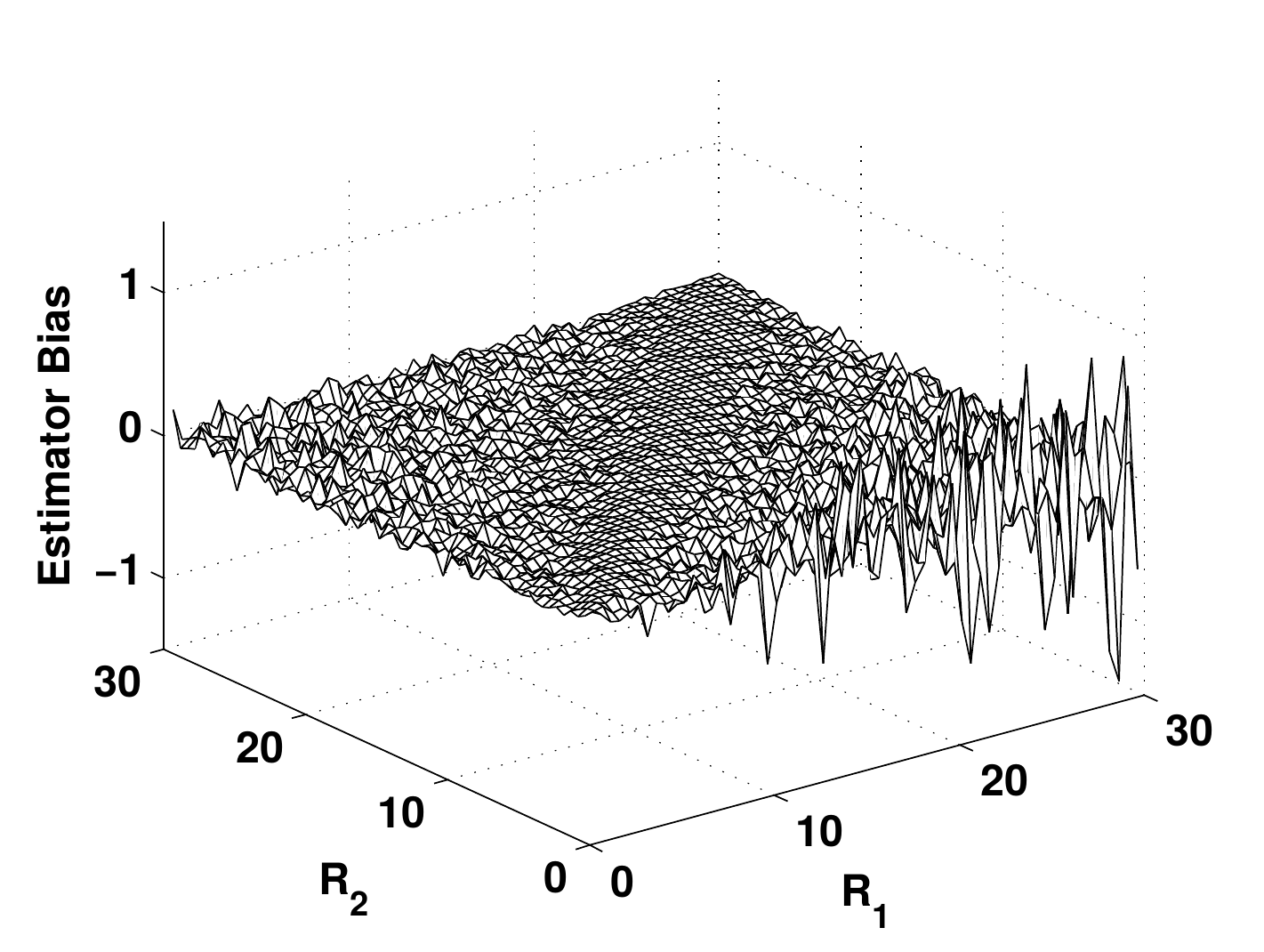}
\caption{Estimator bias for the proposed relative entropy rate estimator. $n$ = 24, $T$ = 250 ms.}
\label{fig:KL Distance Estimator}
\end{figure}

\begin{figure}
\centering
\includegraphics[scale=0.3]{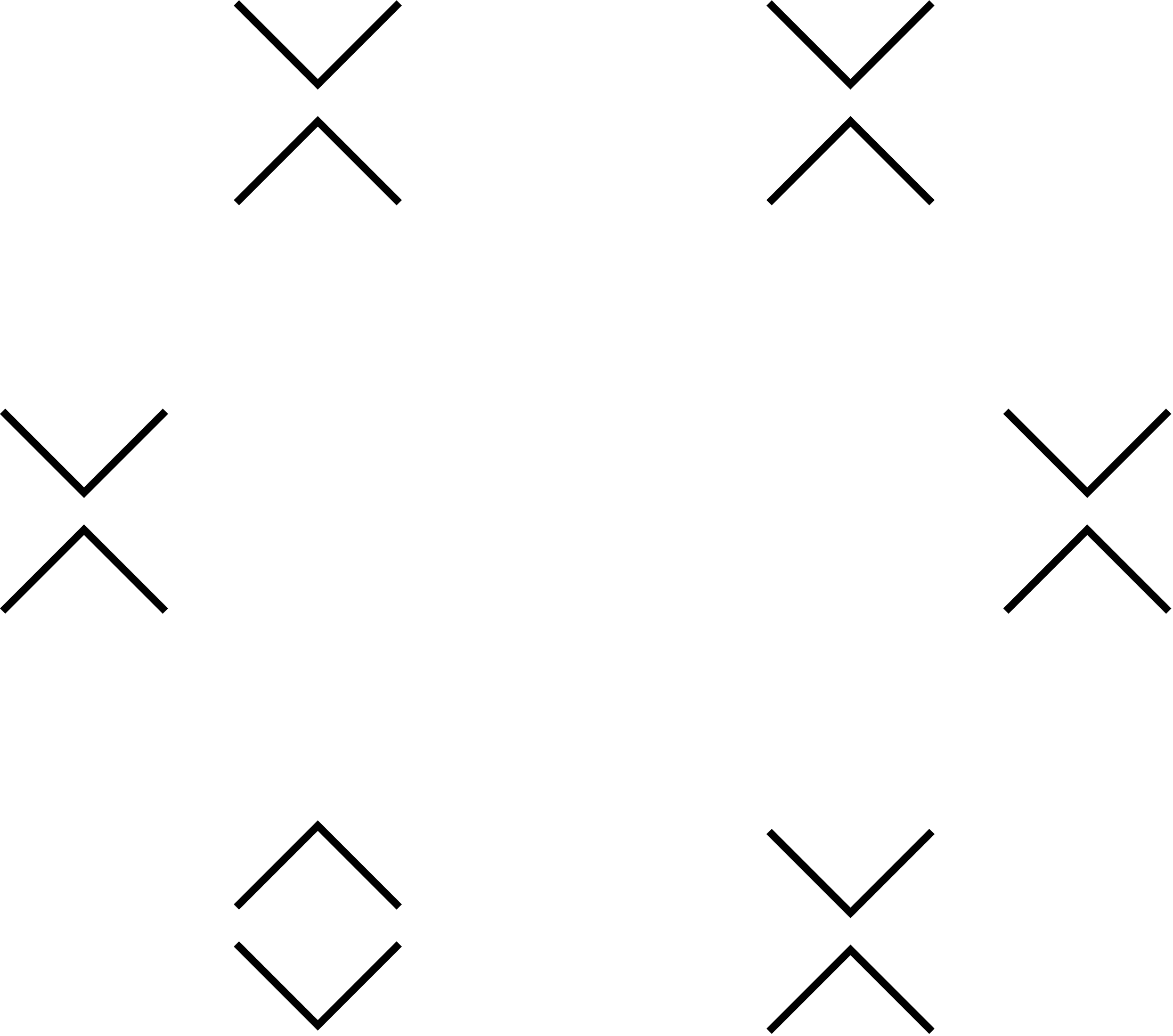}
\caption{Task 1}
\label{fig:fig1a}
\end{figure}

\begin{figure}
\centering
\includegraphics[scale=0.3]{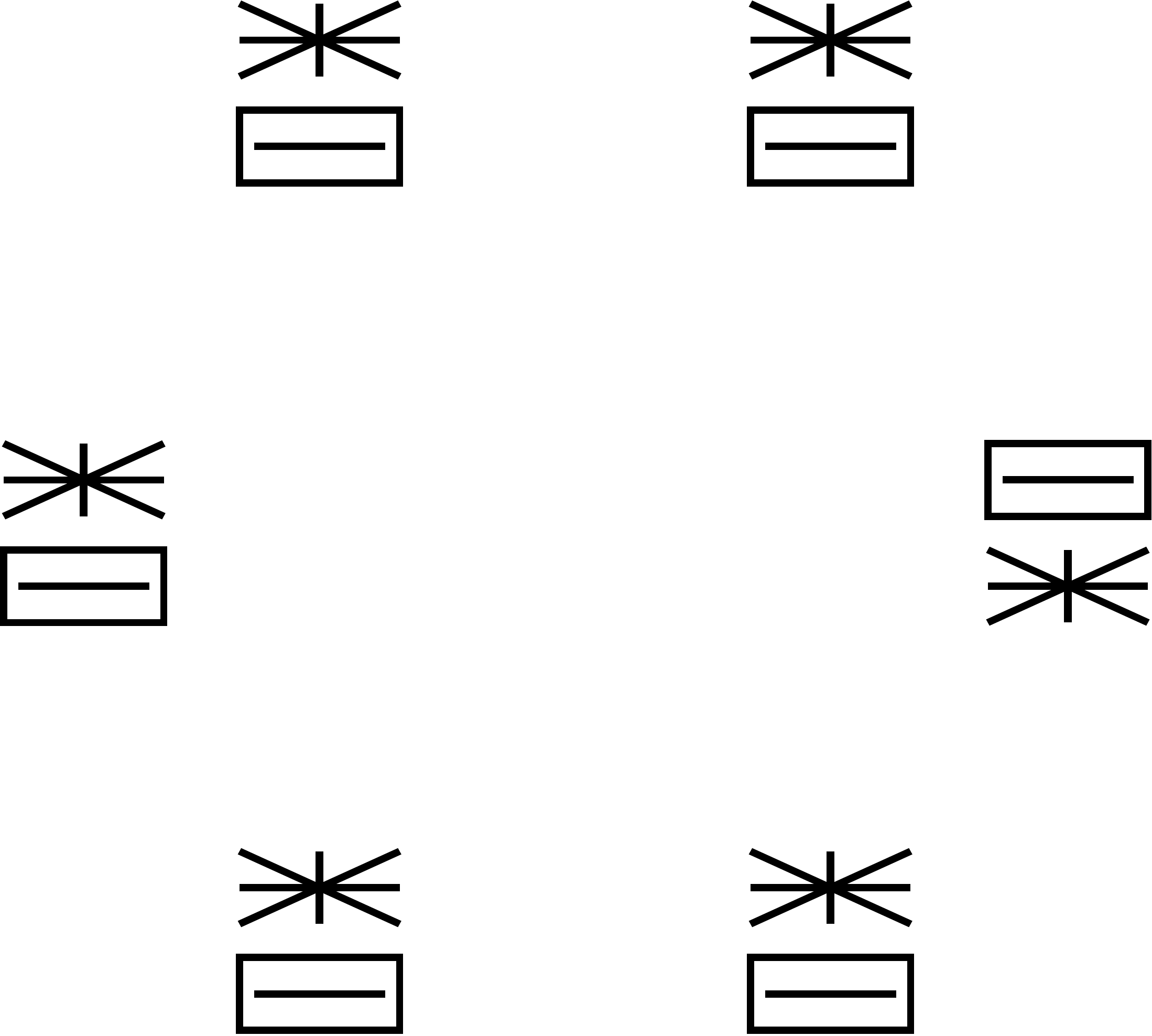}
\caption{Task 2}
\label{fig:fig1b}
\end{figure}

In Figure \ref{fig:KL Distance Estimator} we plot the estimator bias for different $(R_{1}, R_{2})$ pairs for $n$ = 24 and $T$ = 250 ms, motivated by the specific neuronal experimental data of Sripati and Olson \cite{ref:201001JNS_SriOls}. From Figure \ref{fig:KL Distance Estimator} we can see that our proposed estimator has low estimation error for most $(R_{1}, R_{2})$. Estimation error is relatively large only when $R_{1}$ is large and $R_{2}$ is close to zero.

%%%%%%%%%%%%%%%%%%%%%%%%%%%%%%%%%%%%%%%%%%%%%%%%%%%%%%%%%%%%%%%%%%%%%%%%%%%%%%%%%%%%%%%%%%
%%%%%%%%%%%%%%%%%%%%%%%%%%%%%%%%%%%%%%%%%%%%%%%%%%%%%%%%%%%%%%%%%%%%%%%%%%%%%%%%%%%%%%%%%%
\bibliographystyle{../../IEEEtran/bibtex/IEEEtran}
{
\bibliography{../../IEEEtran/bibtex/IEEEabrv,../../BIB/ISITbib}
}

%%%%%%%%%%%%%%%%%%%%%%%%%%%%%%%%%%%%%%%%%%%%%%%%%%%%%%%%%%%%%%%%%%%%%%%%%%%%%%%%%%%%%%%%%%
%%%%%%%%%%%%%%%%%%%%%%%%%%%%%%%%%%%%%%%%%%%%%%%%%%%%%%%%%%%%%%%%%%%%%%%%%%%%%%%%%%%%

\newpage

\end{document}